\documentclass[journal]{IEEEtran}

\usepackage{color}
\usepackage{algorithm,algorithmicx,algcompatible}
\usepackage{verbatim, subfigure}
\usepackage{array, calc, multicol}
\usepackage{xspace}
\usepackage{enumerate}
\usepackage{graphics, epstopdf, graphicx, epsfig, psfrag, epsf, subfigure}
\usepackage{amssymb, amsfonts, amsmath, times}
\usepackage{multicol, multirow, balance, verbatim,caption}
\usepackage[mediumspace,mediumqspace,Grey,squaren]{SIunits}

\usepackage[T1]{fontenc}

\usepackage{url}
\usepackage{breakurl}

\makeatletter
\def\url@leostyle{%
  \@ifundefined{selectfont}{\def\UrlFont{\sf}}{\def\UrlFont{\small\ttfamily}}}
\makeatother
\newcommand{\eg}{{\it e.g.}\xspace}
\newcommand{\ie}{{\it i.e.}\xspace}
\newcommand{\etc}{{\it etc.}\xspace}

\newcommand\textvtt[1]{{\normalfont\fontfamily{cmvtt}\selectfont #1}}
\newcommand{\anteater}{\textvtt{Ant\-Monitor}\xspace}
\newcommand{\anteaterCS}{\textvtt{Ant\-Monitor\-~Client-Server}\xspace}
\newcommand{\anteaterMO}{\textvtt{Ant\-Monitor}\xspace}

\newcommand{\client}{\textvtt{Ant\-Monitor\-~App}\xspace}

\newcommand{\logserver}{\textvtt{Log\-Server}\xspace}
\newcommand{\evaluator}{\textvtt{Ant\-Evaluator}\xspace}
\newcommand{\novpn}{\textvtt{NoVpn}\xspace}
\newcommand{\tPacket}{\textvtt{tPacket\-Capture}\xspace}
\newcommand{\meddle}{\textvtt{Meddle}\xspace}
\newcommand{\swan}{\textvtt{Strong\-Swan}\xspace}
\newcommand{\haystack}{\textvtt{Hay\-stack}\xspace}

\newcommand{\privacyGuard}{\textvtt{Pri\-vacy\-~Guard}\xspace}
\newcommand{\pguard}{\textvtt{Pri\-vacy\-~Guard}\xspace}
\newcommand{\recon}{\textvtt{Recon}\xspace}
\newcommand{\speedtest}{\textvtt{Speedtest}\xspace}

\newfont{\mycrnotice}{ptmr8t at 7pt}
\newfont{\myconfname}{ptmri8t at 7pt}

\begin{document}

%
\title{AntMonitor: A System for On-Device Mobile Network Monitoring and its Applications}

\author{Anastasia Shuba, Anh Le, Emmanouil Alimpertis, Minas Gjoka, Athina Markopoulou
		\thanks{All authors were with UCI when this work was conducted. A. Le is now with Shoelace Wireless and M. Gjoka is with Google.}
		\\
		{\large \{ashuba, anh.le, ealimper, mgjoka, athina\}@uci.edu}
		\\
		{\large University of California, Irvine}
	}

%
%
%

%
%
%
%
%

\maketitle

 \begin{abstract}

 In this paper, we present a  complete system for  on-device passive monitoring, collection, and analysis of fine-grained, large-scale packet measurements from mobile devices. First, we describe the design and implementation of \anteater as  a user-space mobile app based on a VPN-service but only on the device (without the need to route through a remote VPN server) and using only the minimum resources required.  We evaluate our prototype and show that it significantly outperforms prior state-of-the-art approaches: it achieves throughput of over 90 Mbps downlink and 65 Mbps uplink, which is 2x and 8x faster than mobile-only baselines and is $94$\% of the throughput without VPN, while using 2--12x less energy. Second, we show that  \anteater is uniquely positioned to serve as a platform for passive on-device mobile network monitoring and to enable a number of applications, including:  (i) real-time detection and prevention of private information leakage from the device to the network; (ii) passive network performance monitoring; and (iii) application classification and user profiling. We showcase preliminary results from a pilot user study at a university campus.

\end{abstract}

\section{Introduction}

In addition to the large and ever-increasing volume of mobile network traffic \cite{ciscomanos}, %
mobile devices  are used today for a range of personal activities (from communication to financial transactions) and have access to personally identifiable information (PII). User behavior and third-party activity eventually manifest themselves as packets transmitted over the mobile network. Therefore, looking at the network activity on  a mobile device provides a  unique vantage point for monitoring and inferring a wealth of information, including  PII leaks, network performance and user behavior.

There is a rich body of literature %
 \cite{xu2011identifying, chen2012network, falaki2010first, vallina2013rilanalyzer,  kreibich2010netalyzr,  phonelab, sommers2012cell, wei2012profiledroid}  
which studies network traffic traces, and the approaches typically fall into one of two categories: either large-scale but coarse-grained traces obtained in the middle of the network, \ie, traces from Internet Service Providers (ISP) \cite{xu2011identifying,chen2012network}, or fine-grained but small-scale traces from a limited set of users \cite{falaki2010first,vallina2013rilanalyzer}. These limitations, privacy and security concerns, and performance bottlenecks have hindered progress.

In this paper, we present \anteater: a system for on-device passive collection and analysis of fine-grained, large-scale, mobile network measurements. We argue that \anteater is well-positioned to become a high-performance passive monitoring tool for crowdsourcing a range of mobile network measurements, as it combines  the following desired properties: (i) it is easy to install (it does not require administrative privileges) and use (it runs as a service app in the background); (ii) it scales well with the number of users thanks to its on-device design; (iii) it provides users with fine-grained control of which data to monitor or log; (iv) it supports real-time analysis on the device and/or  off-line analysis on a log server, but does not intercept packets in middleboxes  and (iv) it allows for semantic-rich annotation of packet traces with contextual information from the device. 

We start by presenting the design and evaluation of \anteater.  \anteater is a user-space mobile app, based on a VPN service (which intercepts all packets without rooting the phones) and routes packets directly to their destination  (by translating all connections on the device, without redirecting them through a remote VPN server).\footnote{Throughout the paper, we will refer to this approach as {\em mobile-only}, and to using a remote VPN server as {\em client-server}. In both cases, a separate  logging server may be  used for uploading logs from the device for subsequent offline analysis.}
Our software architecture uses the minimum number of resources to achieve the highest performance, including (i) the minimum number of threads for network routing (two: one for reading/writing from/to the TUN) and (ii) the minimum number of sockets (by multiplexing UDP sockets and carefully managing TCP sockets), so as to not interfere with the performance of the foreground apps. 
Thanks to these and other design choices, \anteaterMO significantly outperforms all prior state-of-the-art  approaches, in terms of throughput and energy, without significantly impacting CPU and memory usage.
More specifically, 
experiments show that \anteaterMO achieves 2x (downlink) and 8x (uplink) throughput of state-of-the-art mobile-only
approaches, namely \pguard \cite{song2015privacyguard} and \haystack \cite{razaghpanah2015haystack}.
The achieved downlink throughput is also at 94\% of the throughput without VPN, and almost double the throughput of comparable client-server approaches; all while using 2--12x less energy. We believe that this performance advantage is crucial for the successful adoption of \anteater by users. However much users may care about added-value services,  they are unlikely to install apps that slow down their phones or drain their battery.

Then, we demonstrate that \anteater~naturally lends itself as a platform for a range of applications that build on top of passive on-device network monitoring, which can be of interest to individual users,  network operators, and researchers. First, we use \anteater to detect and prevent in real-time
{\em leakage of private information (PIIs)} from the device to the network. We show that \anteater was able to detect a large volume of PII leaks and the destinations where this information goes to, and we provide  visualization of this information on the mobile device (in real-time) and/or on our \logserver (off-line) \cite{antmonitor-website}. 
Second, we use \anteater for {\em passive performance measurements} network-wide (\eg, network performance maps) as well as per-user (usage profiles) or per-app.   %
This information comes at no additional overhead (\eg, no active probing is needed) since \anteater touches every packet transmitted over the network, and can provide insights into network usage and provisioning as well as fine-grained information.
Third, we use the packet traces collected by \anteater, extract features only from TCP/IP headers, annotate them with rich contextual information available on the device (such as location, time, foreground/background apps, etc.) to train {\em machine learning models}. We use these models for traffic classification of flows to applications (achieving higher classification accuracy than state-of-art methods that use HTTP payloads \cite{meddle}) as well as for user profiling based on minimal information. We present results from a pilot user study at a university campus to demonstrate the capabilities of \anteater and its enabling potential for these applications. 
The structure of the rest of the paper is as follows. Section \ref{sec:related} presents related work. Section \ref{sec:system} describes the objectives, design, and implementation of the \anteater system.
Section \ref{sec:evaluation} presents performance evaluation and comparison
to state-of-the-art approaches. Section \ref{sec:applications} describes the applications of \anteater to  three domains, namely: privacy leak detection and prevention (Section \ref{sec:leakage}); %
passive monitoring of network performance (Section \ref{sec:measurement}); and application and user classification (Section \ref{sec:learning}). Section \ref{sec:conclusion} concludes and discusses directions for future work. %
\section{Related Work} \label{sec:related}

\subsection{VPN-based Mobile Network Monitoring \label{sec:VPN-related}}

The approach we follow in \anteater is passive monitoring on the device, guided by the design objectives outlined in Section \ref{sec:objectives}. 
In particular, the only way to intercept every packet in and out of the mobile device, while running in user space (without root or custom OS), today, is to establish a VPN service on the device.
There are two VPN approaches: client-server and mobile-only, described below.

In {\bf Client-Server VPN} approaches, packets are tunneled from the VPN client on the mobile device to a remote VPN server, where they can be processed or logged. A representative of this approach is \meddle \cite{meddle}, which builds on top of the \swan \cite{swan} VPN software. 
Disadvantages of this approach include the fact that packets are routed through a middle server thus posing additional delay and privacy concerns, lack of client-side annotation (thus no ground truth available at the server), and potentially complex control mechanisms (the client has to communicate the selections of functionalities, \eg, ad blocking, to the server). In \cite{antmonitor15},  a client-server system was proposed that remedied the latter two disadvantages by building a custom client app. An advantage of the client-server approach is that it can be 
combined with other VPN and proxy services (\eg, encryption, private browsing), and be offered by ISPs as an added-value service.

In {\bf Mobile-Only VPN} approaches, the client establishes a VPN service on the phone to intercept all IP packets and does not require a VPN server for routing. It extracts the content of captured outgoing packets and sends them through newly created protected UDP/TCP sockets \cite{androidvpn} to reach Internet hosts; and vice versa for incoming packets. %
This approach may have high overhead due to this layer-3 to layer-4 translation, the need to maintain state per connection and additional processing per packet. If not carefully implemented, this approach can significantly affect network throughput:  for example, see the poor performance of   \tPacket \cite{tpacketcapture} -- an application currently available on Google Play that utilizes this mobile-only approach.  %
 Two state-of-the-art representatives of the mobile-only approach are \haystack \cite{razaghpanah2015haystack,haystack-webpage} and \pguard\cite{song2015privacyguard}. They  both  focus on applying and optimizing their systems for detection of PII  leaks. %
The  \anteaterMO  system describe din this paper, also follows the mobile-only VPN approach but  can achieve 2x and 8x their downlink and uplink throughput, 
as shown  in Section \ref{sec:evaluation} and summarized in Section \ref{sec:ours}.

  \subsection{Other Monitoring Approaches \label{sec:monitoring-related}}

Work on monitoring network traffic generated  by mobile devices can  be roughly classified according to  the vantage point and measurement approach.

 {\bf  Rooted phones and OS approaches.}  
Using a custom OS  or a rooted phone one can get access to fine-grained information on the device, including network traffic.  \textvtt{Phonelab} \cite{phonelab} and others \cite{vallina2013rilanalyzer,falaki2010first, wei2012profiledroid} use packet capture APIs such as \textvtt{tcpdump} or \textvtt{iptables-log}.  %
These are powerful  but  inherently limited to small scale-deployment as the overwhelming majority of  users do not have rooted phones,  and wireless providers and phone manufacturers strongly discourage rooting. The same limitation applies to OS-based approaches, including: \textvtt{TaintDroid} that uses a custom OS, \cite{enck2014taintdroid}, \textvtt{MockDroid} \cite{beresford2011mockdroid}  and \textvtt{AppFence} \cite{hornyack2011thesedroids} that dynamically intercept any permission request to certain resources,
\textvtt{AndroidLeaks} \cite{gibler2012androidleaks} and \textvtt{PiOS} \cite{egele2011pios} that use static analysis.

{\bf Active Measurements from Mobile Devices.}  There are mobile apps, developed  by researchers \textvtt{Netalyzr} \cite{netalyzr-googleplay}, \textvtt{Mobilyzer} \cite{nikMobilyzer:15} or the industry  (\eg, \speedtest, \textvtt{CarrierIQ} or \textvtt{Tutella}),   to perform active network measurements of various metrics (throughput, latency, RSS) from the mobile device. They run in user space, without root, and allow for accurate measurements. However, care must be given to avoid burdening the device's resources and crowdsourcing is often used to distribute the load (see Section 2.3).

{\bf  Passive Monitoring Inside the Network.} ISPs and other organizations sometimes  can capture packets on links in the middle of their networks, \eg at an ISP's  or other organization's network \cite{alcatelNetGuardian:13, cranorGigascope:03,gerber2010speed}. Researchers typically analyze network traces collected by others (\eg large tier-1 networks \cite{xu2011identifying}  or  from university campus WiFi networks \cite{chen2012network}). Limitations of this approach include that (i) it only captures traffic going through the particular measurement point and (ii) it has access only to packet headers (payload is increasingly encrypted), not to ground truth or semantic-rich information (\eg apps that produced the packets). 

\subsection{Applications of Network Monitoring}

\label{sec:privacy-related}

{{\bf Privacy Leaks Detection.}} Next, we review work  on detecting private data leaking out of a device, which is related to our first application  in Section \ref{sec:leakage}. Some approaches require a custom OS or rooting the phone, such as as \textvtt{TaintDroid} \cite{enck2014taintdroid} was one of the early tools built to identify privacy leaks in real-time, and others reviewed in Section \ref{sec:monitoring-related} .

Another approach  is to allow the user to define  strings (\eg, IMEI,  device id, email, or any string corresponding to sensitive information that the user wants to protect) and then monitor for potential leaking of that information from the device to the network. \anteater (as well as \haystack and \pguard)  follow this approach: they monitor, on the device itself, the payload of all outgoing packets, searching for  the predefined strings. In order to detect leaked strings in  encrypted traffic, all three tools need a TLS proxy to first decrypt the traffic before string matching.  Although the goal is the same, implementation matters: \anteater is currently the only tool that can prevent (\ie, block or hash the private string), in addition to detection, on the mobile device without root privileges; \anteater and \pguard can perform real-time detection, while \haystack does not. 

\recon  \cite{recon15}  also inspects packets for privacy leakage, but because it  builds on top of \meddle \cite{meddle},  all packet processing  (including privacy leaks detection) happens not on the device itself but on the Meddle server, with all the advantages and disadvantages of a client-server VPN discussed in Section \ref{sec:VPN-related}. %
 \recon is also the first to use machine learning to identify flows that leak private data without prior knowledge of the users' PII, based on HTTP features and training  on user feedback, as well as on ground truth, manually obtained.  This approach is also applicable to \anteater.

{{\bf  Performance Monitoring.}}
Typically, network performance measurements (\eg signal strength, throughput, delay) are outsourced to companies (some of them drive around and measure the networks) and/or are crowdsourced from individual users.  Examples of {\em third-party companies} include: \textvtt{Carrier IQ} \cite{carrierIQ}, which  is embedded in the firmware of over $150$M smartphones and reports network information, signal strength and the users' location; %
 \textvtt{Tutella}, which  provides a network performance SDK that can be embedded in other mobile applications; %
and \textvtt{Mobilyzer} {\cite{nikMobilyzer:15}}, which provides  an open platform for controllable mobile network measurements. 
Examples of {\em crowdsourcing} projects include \textvtt{Speedtest} \cite{speedtest},  \textvtt{OpenSig-nals} ~\cite{openSignal}, \textvtt{Sensorly} \cite{sensorly}, and \textvtt{RootMetrics}~\cite{rootMetrics}. These companies  make mobile applications that allow users to perform and report active measurements; the companies use that data to release  performance reports \cite{openSignalUSWiFiReport}  and awards \cite{rootMetricsReportsAwards} for cellular and Wi-Fi at points of interest (\eg, metro areas, airports, sports venues, etc). Work in~\cite{sommers2012cell} analyzed crowdsourced data from \speedtest in order to compare the performance of cellular and Wi-Fi networks in large metro areas.  %
 Work in~\cite{deng2014wifi}  released an app for crowdsourcing  measurements of throughput and latency over LTE and Wi-Fi.

Similar to some of these approaches, \anteater crowdsources measurements from an end-user app. Contrary to the above apps, it can  passively infer network performance metrics (e.g. throughput), without sending active probes. It can be used to create both network-wide performance maps and user-specific statistics, as discussed in Section \ref{sec:measurement}.

{{\bf Learning.}} 
 Several  papers \cite{dai2013networkprofiler,miskovic2015appprint, xu2011identifying,yao2015samples} perform {\em app classification} of flows by building signatures  from unencrypted HTTP sessions in network traces: 
 in \cite{xu2011identifying},  the HTTP User-Agent field was used  to map flows into apps; in \cite{dai2013networkprofiler},  HTTP header key-value pairs were used to build unique app signatures that operate on a per-flow basis; in \cite{miskovic2015appprint},  more flows could be identified  by expanding the usage of tokens in the HTTP headers (beyond HTTP request data) and by propagating the identification of a flow mapped to a specific app to other flows that occur at the same time; in \cite{yao2015samples}, conjunctive rules are constructed from HTTP header data.  %
All these methods rely on HTTP headers whereas we perform app classification using only TCP/IP headers in Section \ref{sec:learning}.
 
Early work on behavioral analysis classified protocols based on packet headers: graphlets \cite{karagiannis2005blinc}, profiling tend-hosts  \cite{karagiannis2007profiling}, traffic dispersion graphs  \cite{iliofotou2007network},  subflows \cite{xie2012subflow}. Prior work on {\em user profiling} includes:
\cite{verde2014no}, which uses HMM classifiers on NetFlow records to fingerprint users behind NAT, and \cite{herrmann2010analyzing}, which re-identifies users over different web sessions.

\subsection{\label{sec:ours}Relation to our own prior work}

In our  Sigcomm 2015 C2BID workshop  \cite{antmonitor15}, we presented a preliminary design and evaluation of \anteater based on the client-server VPN approach.  In Mobicom 2015 Demos (3-page paper and best demo in S3)  \cite{antmonitor-poster-mobicom15}, we also  demonstrated the use of that client-server version to detect privacy leaks. In contrast, in this paper, we present the mobile-only design of \anteater and we show that it significantly outperforms all state-of-the-art approaches, including previous mobile-only \haystack \cite{razaghpanah2015haystack,haystack-webpage} (app available since Oct. 2015) and \pguard \cite{song2015privacyguard} (open source), as well as server-based approaches (\eg, \swan, \meddle \cite{meddle}, and our own prior \anteater client-server system), as discussed in Section \ref{sec:evaluation}.  In this paper we also present three possible applications of the \anteaterMO system, namely privacy leaks (Section \ref{sec:leakage}), performance monitoring (Section \ref{sec:measurement}), and learning of network-level behavior (Section \ref{sec:learning}), based on an 8 month user study (which is analyzed for the first time in this paper). We have outlined the first and the second applications in  2-page posters  \cite{antmonitor-poster-mobicom15} and \cite{nsdi-poster-17}, respectively, while the third application is presented for the first time in this paper.

\section{The AntMonitor System} \label{sec:system}
\subsection{Design Rationale}
 \label{sec:objectives}
Here we describe the main objectives of \anteater and the key design choices made to meet the objectives. %

{\em Objective 1:  Large-Scale Measurements:} \anteater is intended for crowdsourcing data from a large number of users, which poses a number of system requirements. First, the app on the mobile device must run without administrative privileges (root access).
To that end, we use the public Virtual Private Network (VPN) API \cite{androidvpn} provided by the Android OS (version 4.0+), which runs on more than 95\% of Android devices \cite{androidpie}.
 Second, in order for a large number of mobile users to adopt it, user experience must not be affected: monitoring must occur seamlessly in the background while the user continues to use the mobile device as usual, and the overhead on the device must be negligible in terms of network throughput, CPU, battery, and data cost. Third, the performance  must scale with the number of users, which makes a strong case for a mobile-only design.

{\em Objective 2: Making it Attractive for Users:}
There must also be incentives for users to participate and \anteater is designed with the capability to offer users a variety of services. The current prototype  offers enhanced privacy protection  (\eg, preventing leakage of private information) and visualizations that help users understand where their data flows (\eg see Fig. \ref{fig:screenshots}(f)). Additional services can be implemented completely on the mobile, such as enhanced wireless network performance by switching among available networks; see Sec. \ref{sec:measurement}.
The \client is also designed to provide users with control over which data they choose to monitor and log, including which applications to monitor, and whether to log full packets or headers only.

{\em Objective 3: Fine-Grained Information:}
 \anteater supports full-packet capture of incoming and outgoing traffic
 in PCAP Next Generation format \cite{pcapng}, which allows to append arbitrary information alongside the raw packets. This is important because, in many cases, this contextual information may only be collected accurately on the mobile  at the time of packet capture, and can play a critical role in subsequent analyses. Examples of such contextual information include names of apps that generate the packets (thus providing  ground truth for application classification, see Section \ref{sec:learning}), location, background apps, and information about the network used (\eg WiFi or LTE, network speed, \etc).

\begin{figure}[t!]
	\centering
	\includegraphics[height=4.35cm]{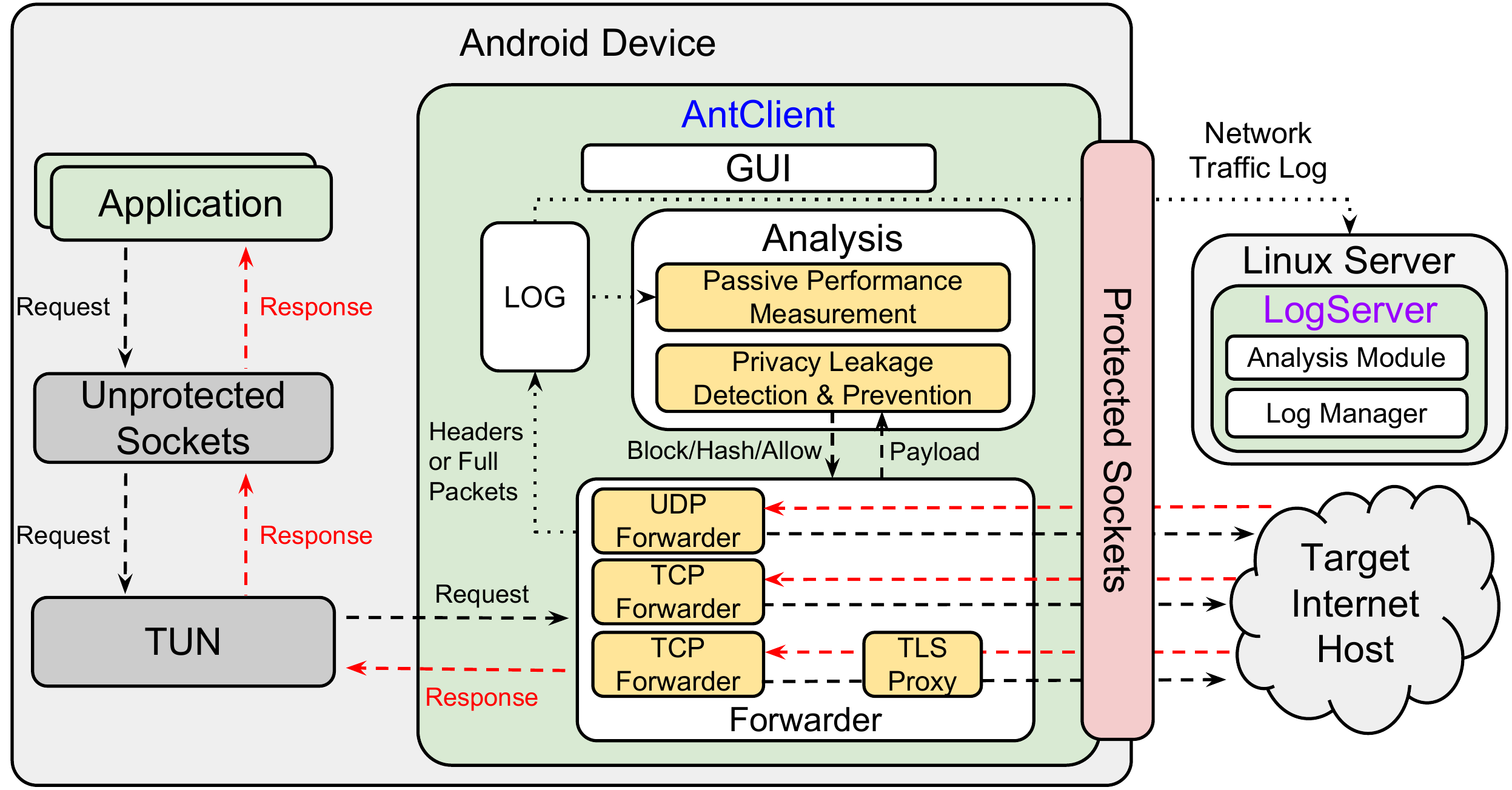}
	\caption{The \anteater Architecture. It consists of  the \client (on the device) and a remote \logserver.}\label{fig:client-based}
\end{figure}

\subsection{System Design}
To support the above design objectives, \anteater is designed to provide the following four main functionalities: traffic interception, routing, logging, and analysis. The rationale for the first two is described here and implementation and optimizations for all four are provided in Sec. \ref{sec:system-implementation} and \ref{sec:opt}. 

{{\bf Traffic Interception.}} The \client establishes a VPN service on the device that runs seamlessly in the background. The service is able to intercept all outgoing and incoming IP datagrams  by creating a virtual (layer-3) TUN interface \cite{androidvpn} and updating the routing table so that all outgoing traffic, generated by any app on the device, is sent to the TUN interface. 
The \client then routes the datagrams to their target hosts on the Internet (as described below). When a host responds, the response will be routed back to the \client, which then sends the response packets to the apps by writing them to TUN. 

{{\bf Traffic Routing.}} To route IP datagrams generated by the mobile apps and arriving at the TUN interface, the intuitive solution would be to use raw sockets, which unfortunately is not available on non-rooted devices. Therefore, the datagrams have to be sent out using layer-4 (UDP/TCP) sockets, which can be done in one of the following two ways:

{\em 1. Client-Server Routing:} This follows the design of a typical VPN service: all traffic is routed through a VPN server \cite{swan, meddle, antmonitor15}.
The main advantage is the simplicity of implementation: the routing is done seamlessly by the operating system at the server with IP forwarding enabled. However, in a crowdsourcing system with a large number of users, sending all traffic through a VPN server faces scalability challenges. Furthermore, users may not want their traffic to be redirected. Therefore, we used an alternative routing approach that can work entirely on the mobile device, without the need of a VPN server, as described next.

{\em 2. Mobile-Only Routing:} Routing IP datagrams to target hosts through layer-4 sockets requires a {\em translation between layer-3 datagrams and layer-4 packets}. For outgoing traffic, data of the IP datagrams has to be extracted and sent directly to the target hosts through UDP/TCP sockets. When a target host responds, its response data is read from the UDP/TCP sockets and must be wrapped in IP datagrams, which are then written to the TUN interface. 
The Mobile-Only design removes the dependency on a VPN server, thus making \anteater  self-contained and  easy to scale. Furthermore, this design enhances user privacy as all data can now stay on the mobile device and is not routed through a middlebox. Only when the user opts in, select packets are temporarily logged on the \client, and then uploaded to the remote \logserver, as described in the next section. 

\subsection{System Implementation}
\label{sec:system-implementation}

The \anteater system architecture is shown in Fig. \ref{fig:client-based}: it consists of  the \client and a \logserver. 

\subsubsection{On the Mobile: \client}

\begin{figure*}[t!]
	\begin{center}
		\subfigure[Home Screen]{\includegraphics[width=2.7cm]{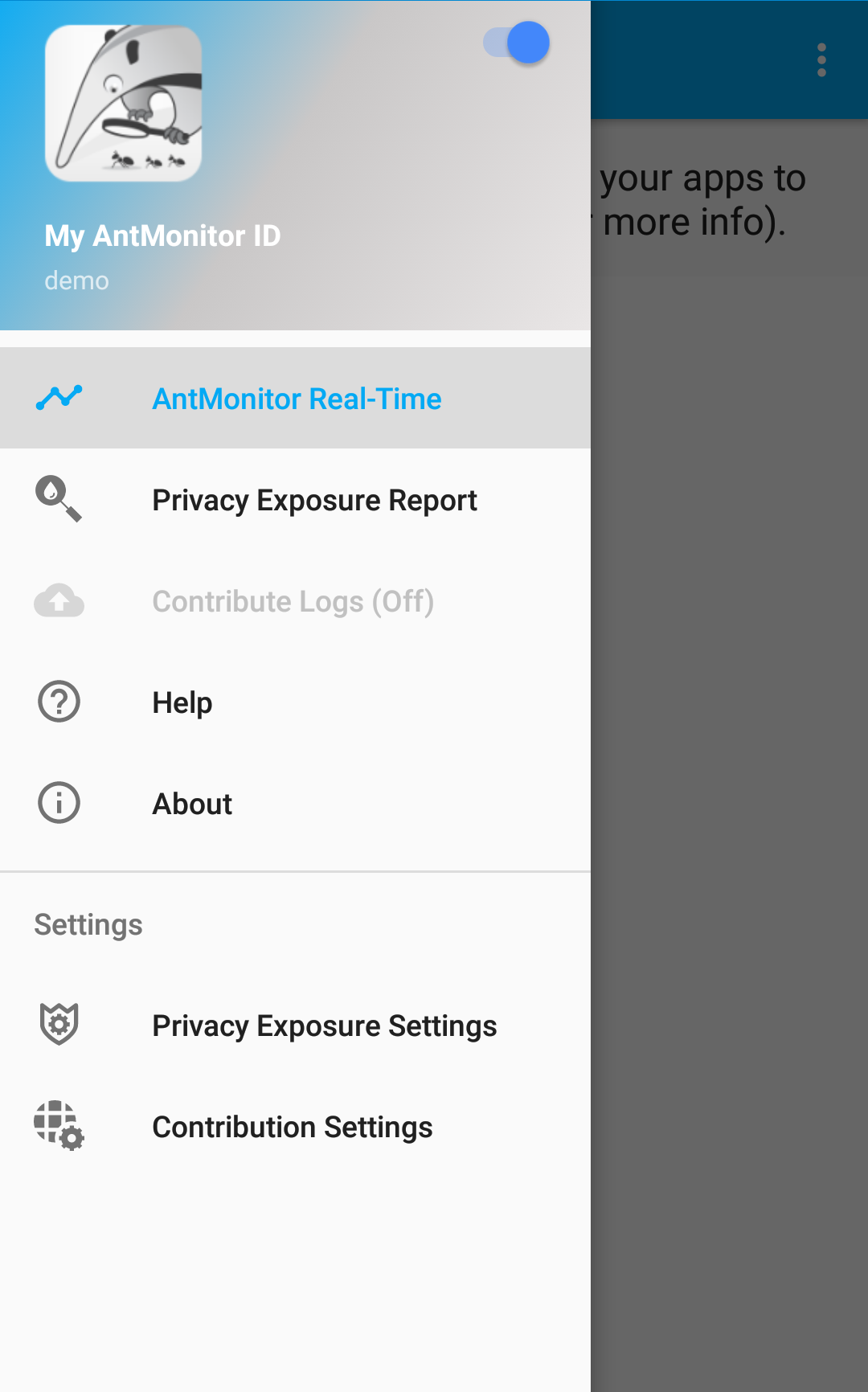}\label{fig:guiMain}\label{fig:guiON}\hspace{2.5pt}}
		\subfigure[Apps to Monitor]{\hspace{2.5pt}\includegraphics[width=2.7cm]{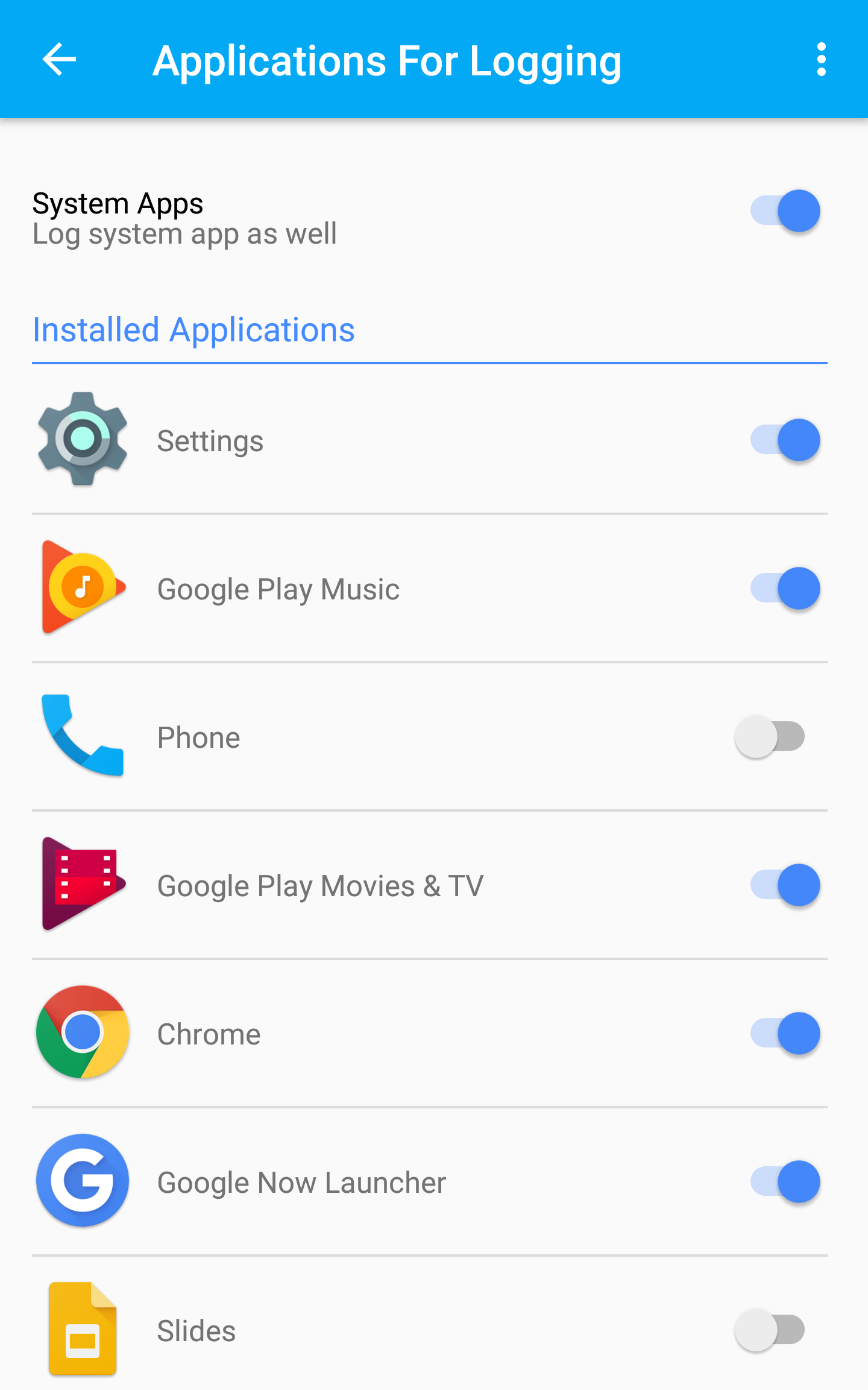}\label{fig:guiSelect}\hspace{2.5pt}}
		\subfigure[Strings to Inspect]{\hspace{2.5pt}\includegraphics[width=2.7cm]{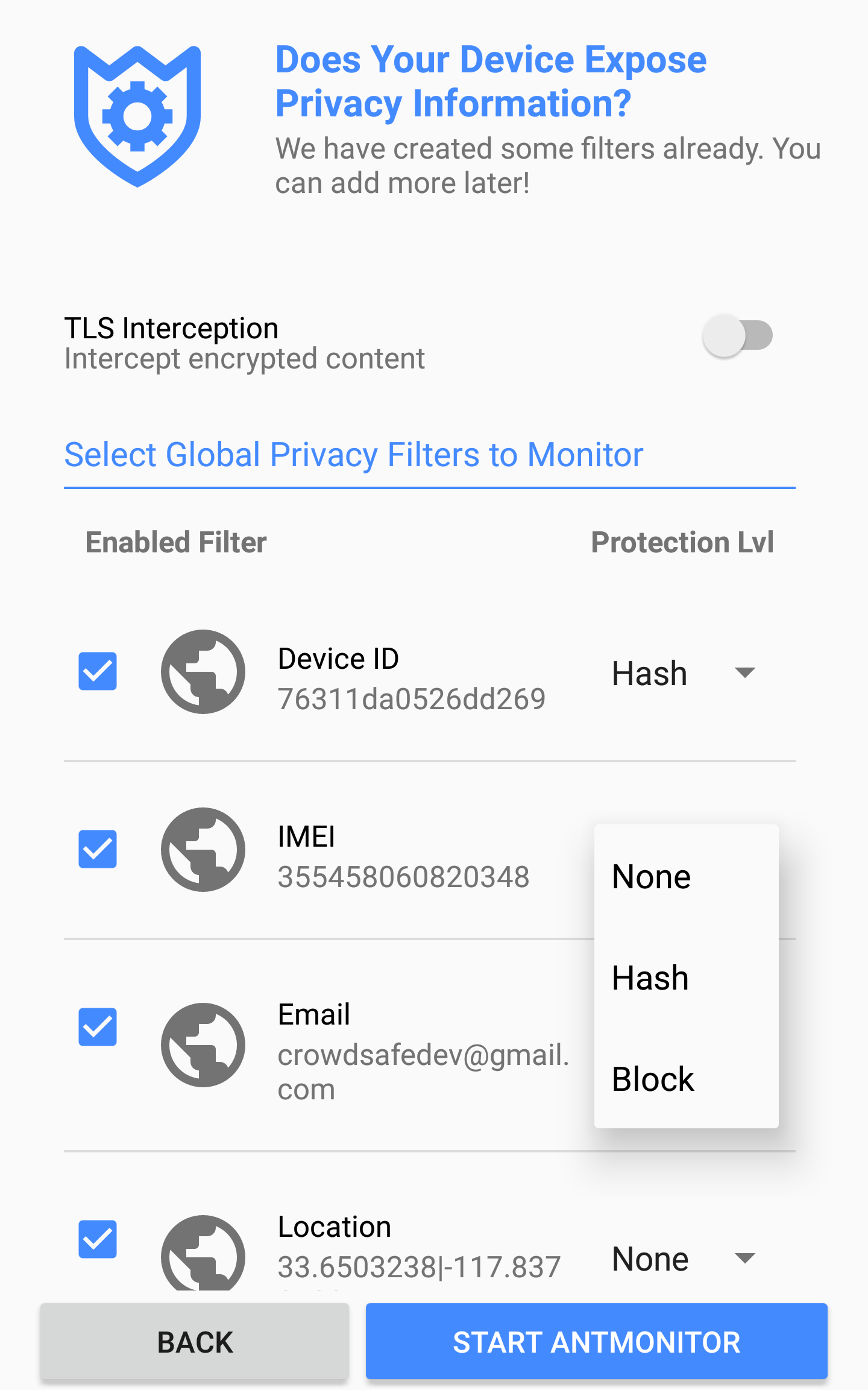}\label{fig:guiPI}\hspace{2.5pt}}
		\subfigure[Privacy Leak Alert]{\hspace{2.5pt}\includegraphics[width=2.7cm]{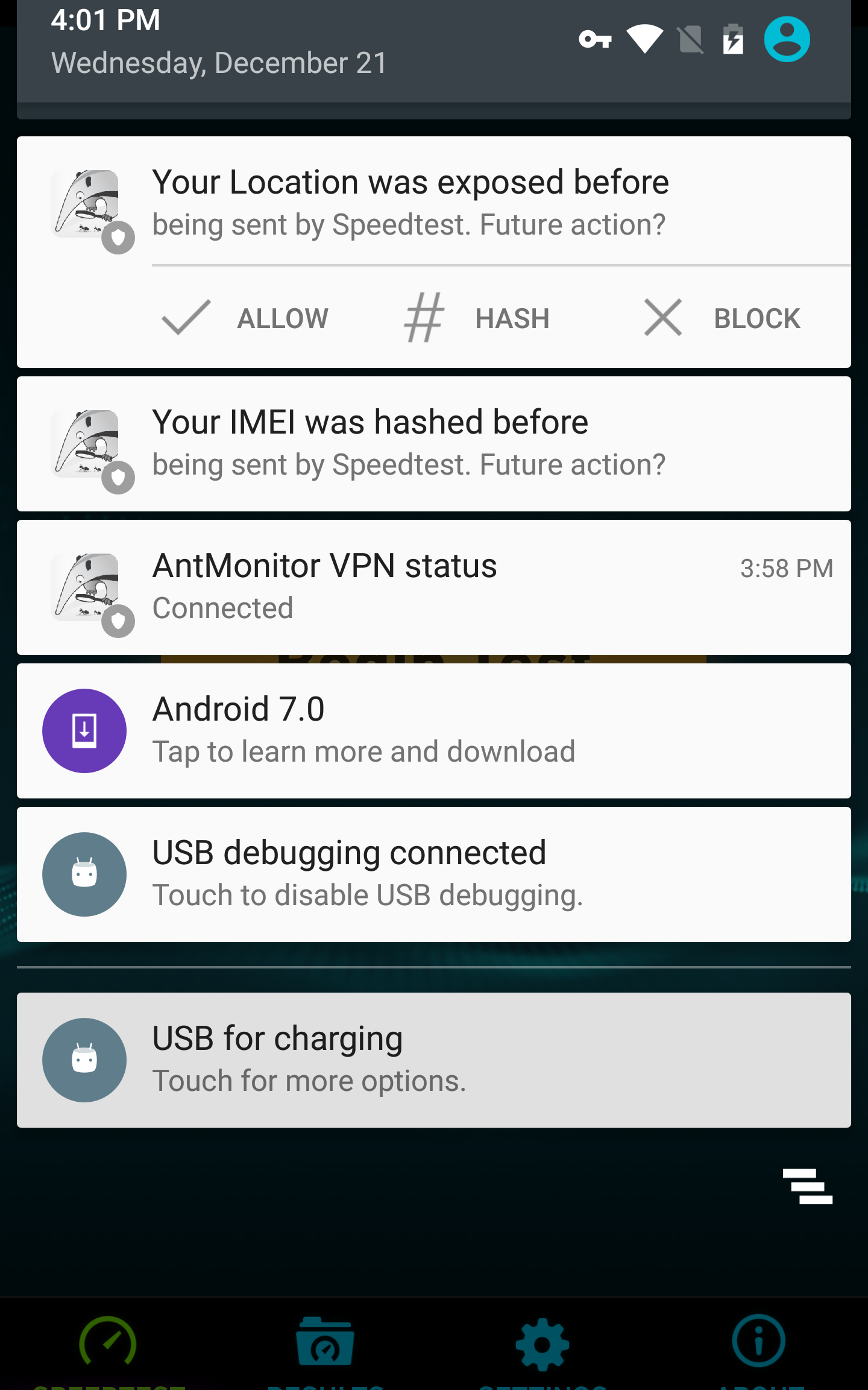}\label{fig:guiNotif}\hspace{2.5pt}}
		\subfigure[History of Leaks]{\hspace{2.5pt}\includegraphics[width=2.7cm]{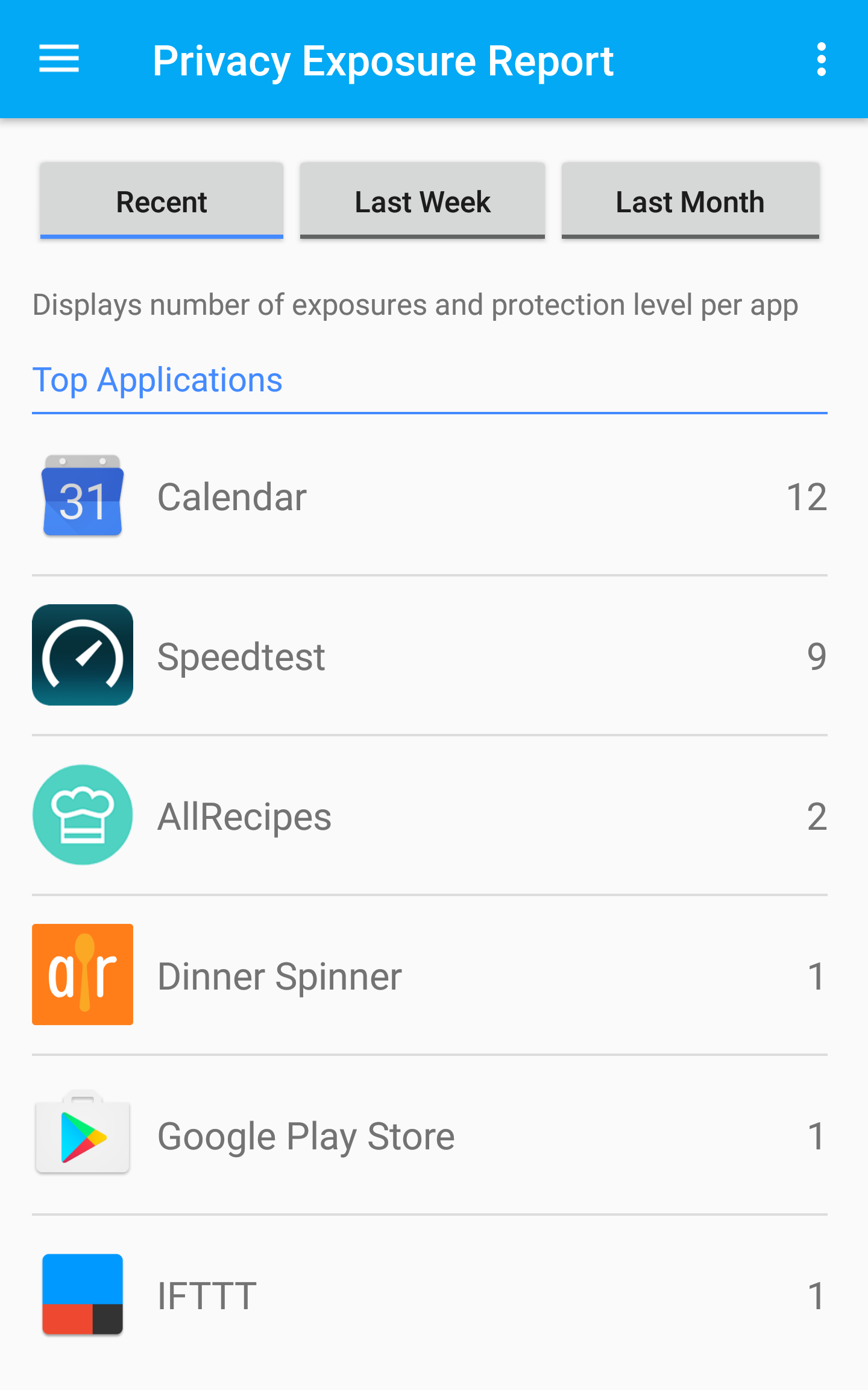}\label{fig:guiHistory}\hspace{2.5pt}}
		\subfigure[Visualization]{\hspace{2.5pt}\includegraphics[width=2.7cm]{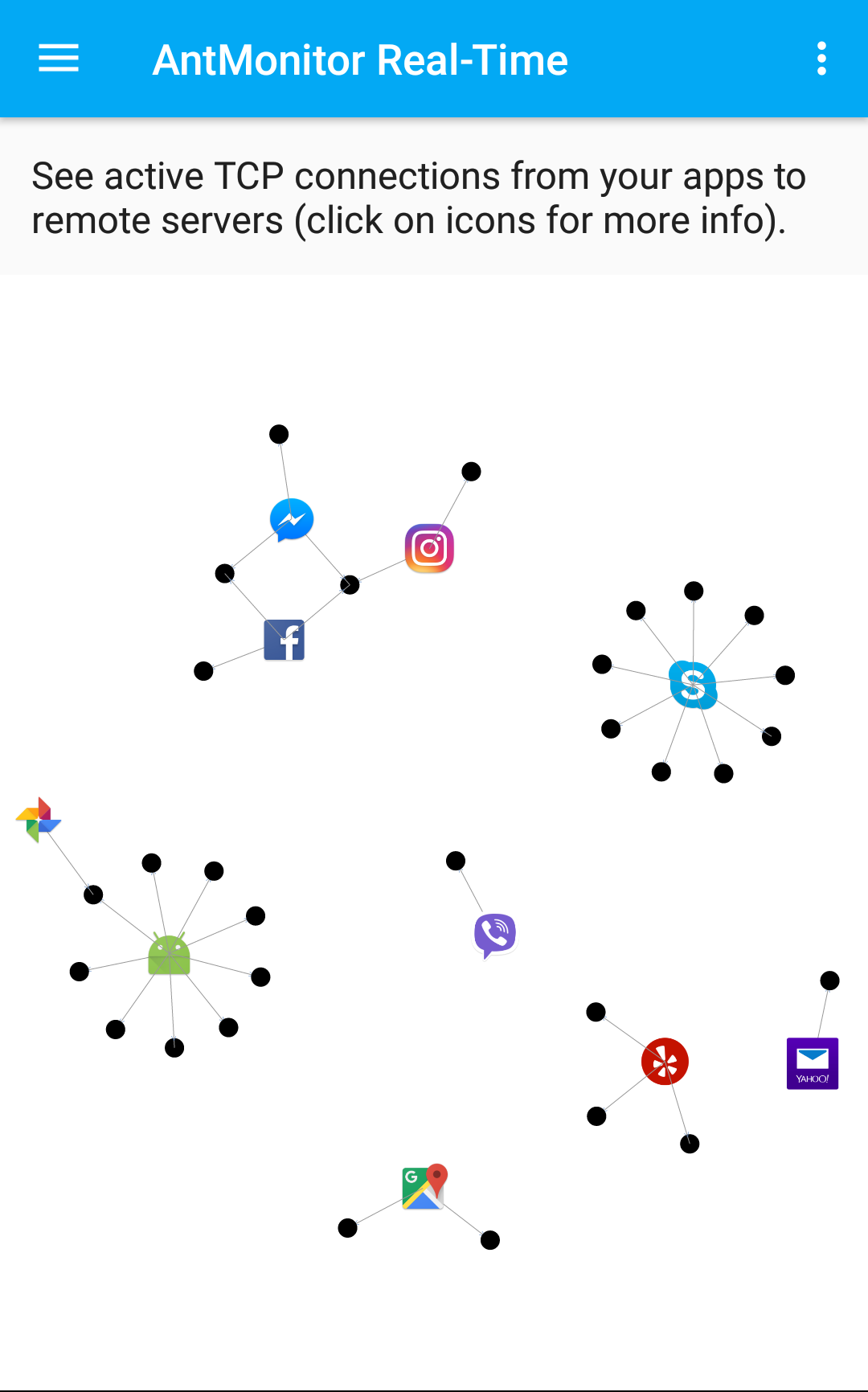}\label{fig:guiVisual}}
		\caption{Screenshots of the \anteater prototype v0.1.5. (a)(b) apply to all uses of the app. (c-f) are specific to Privacy Leaks.  (f) shows which apps send packets  to which IP destinations for all active connections and is updated in real-time.}
		\label{fig:screenshots}
	\end{center}   
\end{figure*}

Our current prototype is an Android app\footnote{It relies on the VPN service which is available on 95\%, \ie billions,  of Android devices today. The VPN API has also just been released on iOS version 9.0+ in Sep. 2015; thus, this approach can  be implemented on iOS, although our prototype is in android.}. In addition to traffic interception and routing, the app contains a Graphical User Interface (GUI)  and modules for logging (Log) and real-time and/or offline Analysis of packets (see Fig. \ref{fig:client-based}).

Fig. \ref{fig:screenshots} shows screenshots of \client's {\bf GUI} for  the AntMonitor App v0.1.5, as available in open beta-testing on Google Play \cite{antmonitor-app}. It allows the user to turn the background VPN service on and off Fig. \ref{fig:guiON}, and to
select which applications should be monitored and logged  as in Fig. \ref{fig:guiSelect} (currently disabled on Google Play \cite{antmonitor-app}, but internally it can allow users to contribute full packets or headers only). Fig. \ref{fig:guiVisual} displays to which destination (\eg servers) each app sends data as a graph of connections updated in real-time. Fig. \ref{fig:guiPI}, \ref{fig:guiNotif} and \ref{fig:guiHistory} are related to the application of privacy leakage detection described in Section \ref{sec:leakage}.

{The \bf Forwarder} manages the  TUN interface and routes network traffic. It consists of two main components: UDP and TCP Forwarder, both depicted in Fig. \ref{fig:client-based}.

The {\em UDP Forwarder} is  simpler, since UDP connections are stateless. When an app sends out an IP datagram containing a UDP packet, the UDP Forwarder records the mapping of the source and destination tuples (a tuple consists of an IP address and a port number), to be used later for reverse lookup. The Forwarder then extracts the data of the UDP packet and sends the data to the remote host through a {\em protected} UDP socket. When a response is read from the UDP socket, the Forwarder creates a new IP datagram,  changes the destination tuple to the corresponding source tuple in the recorded mapping, and writes the datagram to TUN.

The {\em TCP Forwarder} works like a proxy server. For each TCP connection made by an app on the device, a TCP Forwarder instance is created. This instance maintains the TCP connection with the app by responding to IP datagrams read from the TUN interface with appropriately constructed IP datagrams. This entails following the states of a TCP connection (LISTEN, SYN\_RECEIVED, ESTABLISHED, \etc) on both sides (app and TCP Forwarder) and careful construction of TCP packets with appropriate flags (SYN, ACK, RST, \etc), options, and sequence and acknowledgment numbers. At the same time, the TCP Forwarder creates an external TCP connection to the intended remote host through a {\em protected} socket to forward the data that the app sent to the server and the response data from the server to the app.

{The \bf Analysis Module} can do both offline and online analysis on intercepted packets. The online capability allows it to take action on live traffic, \eg, preventing private information from leaking. Since the analyses are done on the mobile, private information is never leaked out of the device, setting \anteater apart from systems like \meddle \cite{meddle}, that perform leakage analysis at a VPN server. In order to inspect encrypted traffic, we implement a TLS proxy: see Sec. \ref{sec:privacy-design}. 

{The \bf Log Module} writes packets to log files on the device, subject to the user preferences of what apps and what information (\eg entire packets or just packet headers) to log.  This module can add rich contextual information to the captured packets by using the PCAP Next Generation format \cite{pcapng}; \eg we currently store application names and network statistics (discussed in detail in Sec. \ref{sec:measurement}) alongside the raw packets. The mapping to app names is done by looking up the packets' source and destination IPs and port numbers in the list of active connections available in $\mathsf{/proc/net}$, which provides the UIDs of apps responsible for each connection. Given a UID, we get the corresponding package name using Android APIs. Finally, the Log Module periodically uploads the log files to \logserver over HTTPS, when the device is charging and is connected to Wi-Fi or upon user's request.

\subsubsection{Data Collection Server: \logserver}

{The \bf Log Manager} supports uploading of files using multi-part content-type HTTPS. For each uploaded file, it checks if the file is in proper PCAPNG format. If so, for each mobile, the manager stores all of its files in a separate folder.

{The \bf Analysis Module} extracts features from the log files and inserts them into a MySQL database to support various types of analyses.  Compared to the Analysis Module of the \client, this module (on the \logserver) has access to crowdsourced data from a large number of devices, and can perform global large-scale analyses. For instance, it could detect global threats and outbreaks of malicious traffic.

\subsection{Novelty of Design\label{sec:novelty}}

The key novelty of our design is that we are able to use the minimum number of resources to achieve highest performance. Specifically, we use the minimum number of: (i) threads (two) for network routing, and (ii) sockets. 

{\bf Thread Allocation.} We have fully utilized Java {New I/O (NIO)} with non-blocking sockets for the implementation of the Forwarder. In particular, Forwarder is implemented as a high-performance (proxy) server, that is capable of serving hundreds of TCP connections (made by the apps) at once, while using only two threads: one thread is for reading IP datagrams from the TUN and another thread is for actual network I/O using the Java NIO Selector and for writing to TUN. Minimizing the number of threads used is critical on a resource constrained mobile platform so that \anteater (which runs in the background) does not interfere with other apps that run in the foreground. For comparison, \privacyGuard creates one thread per TCP connection, which rapidly exhausts the system resources even in a benign scenario, \eg, opening the CNN.com page could create about 50 TCP connections, which results in low performance (see Sec. \ref{sec:evaluation}). %

{\bf Socket Allocation.} Since the Forwarder needs to create sockets to forward data and the Android system imposes a limit of 1024 open file descriptors per user process, sockets must be carefully managed. To this end, we minimize the number of sockets used by the Forwarder by (i) multiplexing the use of UDP sockets: we use a single UDP socket for all UDP connections, and (ii) carefully managing the life cycle of a TCP socket to reclaim it as soon as the server or the client closes the connection. For comparison, \privacyGuard uses 1 socket per UDP connection and 2 sockets per TCP connection; \haystack uses 1 socket per UDP connection and 1 socket per TCP connection.

\subsection{Performance Optimization\label{sec:opt}}

\begin{figure}[t!]
\centering
\includegraphics[height=1.8cm]{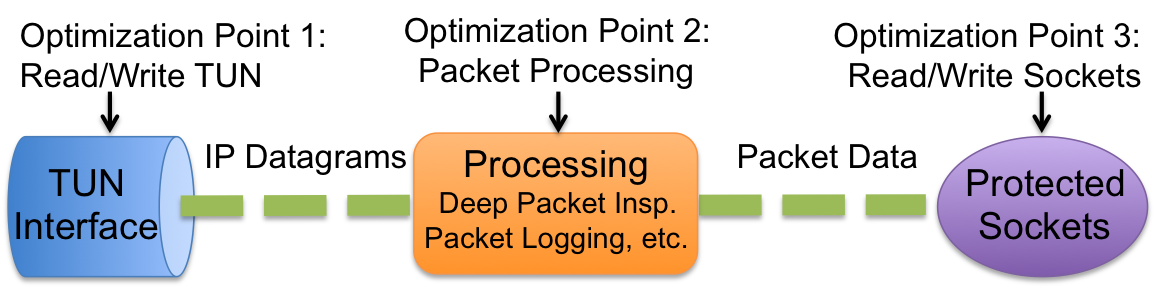}
\caption{{Performance Optimization.} \label{fig:optimization}}
\end{figure}

Since the \client processes raw IP datagrams in  user-space, it is highly non-trivial to achieve good performance. We  investigated the performance bottlenecks of our approach specifically and VPN approaches in general. These bottlenecks are depicted in Fig. \ref{fig:optimization}. We then address these bottlenecks through a combination of  techniques, from typical optimization techniques (including implementing custom native C libraries and deploying high-performance network I/O patterns) to highly customized techniques that we specifically devised for the VPN-based architecture. We also provide a detailed comparison of our design to \privacyGuard \cite{song2015privacyguard} (whose source code is publicly available); in contrast,  \haystack's source code is unavailable, therefore we qualitatively compare a subset of  techniques that we could infer from \haystack's  description \cite{razaghpanah2015haystack} and  our observations.

{\bf Traffic Routing (Points 1, 2, and 3).} The techniques that we adopted are as follows: (i) we explicitly manage and utilize a Direct ByteBuffer for I/O operations with the TUN interface and the sockets, (ii) we store packet data in byte arrays, and (iii) we minimize the number of copy operations and any operations that traverse through the data byte-by-byte. These techniques are based on the following observations: Direct ByteBuffer gives the best I/O performance because it eliminates copy operations when I/O is performed in native code. Plus, Direct ByteBuffer on Android is actually backed by an array (which is not typically the case on a general Linux platform); therefore, it creates synergy with byte arrays: making a copy of the buffer to a byte array (for manipulation) can be done efficiently by performing a memory block copy as opposed to iterating through the buffer byte-by-byte. (Memory copy is also used whenever a copy of the data is needed, \eg, for IP datagram construction.) 
Finally, because the allocation of a Direct ByteBuffer is an expensive operation, we carefully manage its life cycle: for an I/O operation, \ie, read from TUN, we reuse the buffer for every operation instead of allocating a new one.

{\bf TUN Read/Write (Point 1).} The Android 
API does not provide a way to {\em poll} the TUN interface for available data. The official Android tutorial \cite{toyvpn}, as well as \privacyGuard and \haystack \cite{song2015privacyguard, razaghpanah2015haystack}, employ periodic sleeping (\eg, 100 ms) between read attempts. This results in wasted CPU cycles if sleeping time is small, or in slow read speed if the sleeping time is large, as the data may be available more frequently than the sleep time. To address this issue, we implemented a {\em native C library} that performs the native {\em poll()} to read data to a Direct ByteBuffer (which is then available in the Java code without extra copies).

It is also important  to read from (and write to) the TUN interface in large blocks to avoid the high overhead of crossing the Java-Native boundary and of the system calls (read() and write()). In early implementations, we observed that IP datagrams read from the TUN interface have a maximum size of 576 B (which is the minimal IPv4 datagram size). This results in maximum read speed of about 25 Mbps on a Nexus 6 for a TCP connection, thus limiting the upload speed. We were able to increase the datagram size by (i) increasing the MTU of the TUN interface to a large value, \eg, 16 KB and (ii) including an appropriate Maximum Segment Size (MSS) in the TCP Options field of SYN-ACKs sent by TCP Forwarder when responding to apps' SYN datagrams. These changes help to ensure that an app can acquire a high MTU when performing Path MTU Discovery, so that each read from TUN results in a large IP datagram. This results in the maximum read speed, \ie, more than {80} Mbps on our Nexus 6. Similarly, it is  important to write to TUN in large blocks: we construct large IP datagrams to write to TUN. We have experimented with other large block values (e.g., 8K, 32K) and found that 16 KB achieved the highest throughput on Nexus 6.

{\bf Socket Read/Write (Point 3).} Similar to when interacting with the TUN interface, in order to achieve high throughput, it is important to read from (and write to) TCP sockets in large blocks. In particular,
we match the size of the buffer used for socket read (\eg, 16 KB minus 40 B for TCP and IP headers) to the size of the buffer used for TUN write (\eg, 16 KB). Similarly, we also matched the size of the buffer used for socket write to that of the buffer used for TUN read. Thus, sending a large IP datagram read from TUN might involve sending several smaller IP datagrams (from the TCP network socket), \ie, fragmentation; however, this is done efficiently in the kernel space. For comparison, \privacyGuard does not implement this matching. 

{\bf Packet-App Mapping (Point 2).} Android keeps active network connections in four separate files in the \textvtt{/proc/net} directory: 
one each for UDP, TCP via IPv4 and IPv6.
Because parsing these files is an expensive I/O operation, we implemented the reading and parsing  in a native C library. Furthermore, to minimize the number of times we have to read and parse them, we store the mapping of app names to source/destination IPs and port numbers in a Hash Map. When the Log Module receives a new packet, it first checks the Map for the given IP and port number pair. If the mapping does not exist, the Log Module re-parses the $\mathsf{/proc}$ files and updates the Map.  \privacyGuard and \haystack also provide packet-app mapping, but their implementation is in Java, and \privacyGuard does not provide logging.

{\bf DPI: Deep Packet Inspection (Point 2).} Although inspecting every packet is costly, we leverage the Aho-Corasick algorithm \cite{multi} written in native C to perform real-time detection without significantly impacting throughput and resource usage (see Sec. \ref{sec:Impact of Log+DPI}). However, this alone is not enough: we must also minimize the number of copies  of each packet. Although the algorithm generally operates on Strings, \client uses Direct ByteBuffers for  routing, and creating a String out of a ByteBuffer object costs us one extra copy. Moreover, Java Strings use UTF-16 encoding and JNI Strings are in Modified UTF-8 format. Therefore, any String passed from Java to native C requires another copy while converting from UTF-16 to UTF-8 \cite{jniStrings}. To avoid two extra copies, we pass the Direct ByteBuffer object and let the Aho-Corasick algorithm interpret the bytes in memory as characters. This  enables us to perform an order of magnitude faster than Java-based approaches (Sec. \ref{sec:Other Metrics}), \ie, compared to \privacyGuard's Java-based string matching and \haystack's Java-based Aho-Corasick implementation.

 \section{Performance Evaluation}
\label{sec:evaluation}

\begin{figure*} [t!]
	\centering
	\subfigure[Download]{\includegraphics[height=5cm]{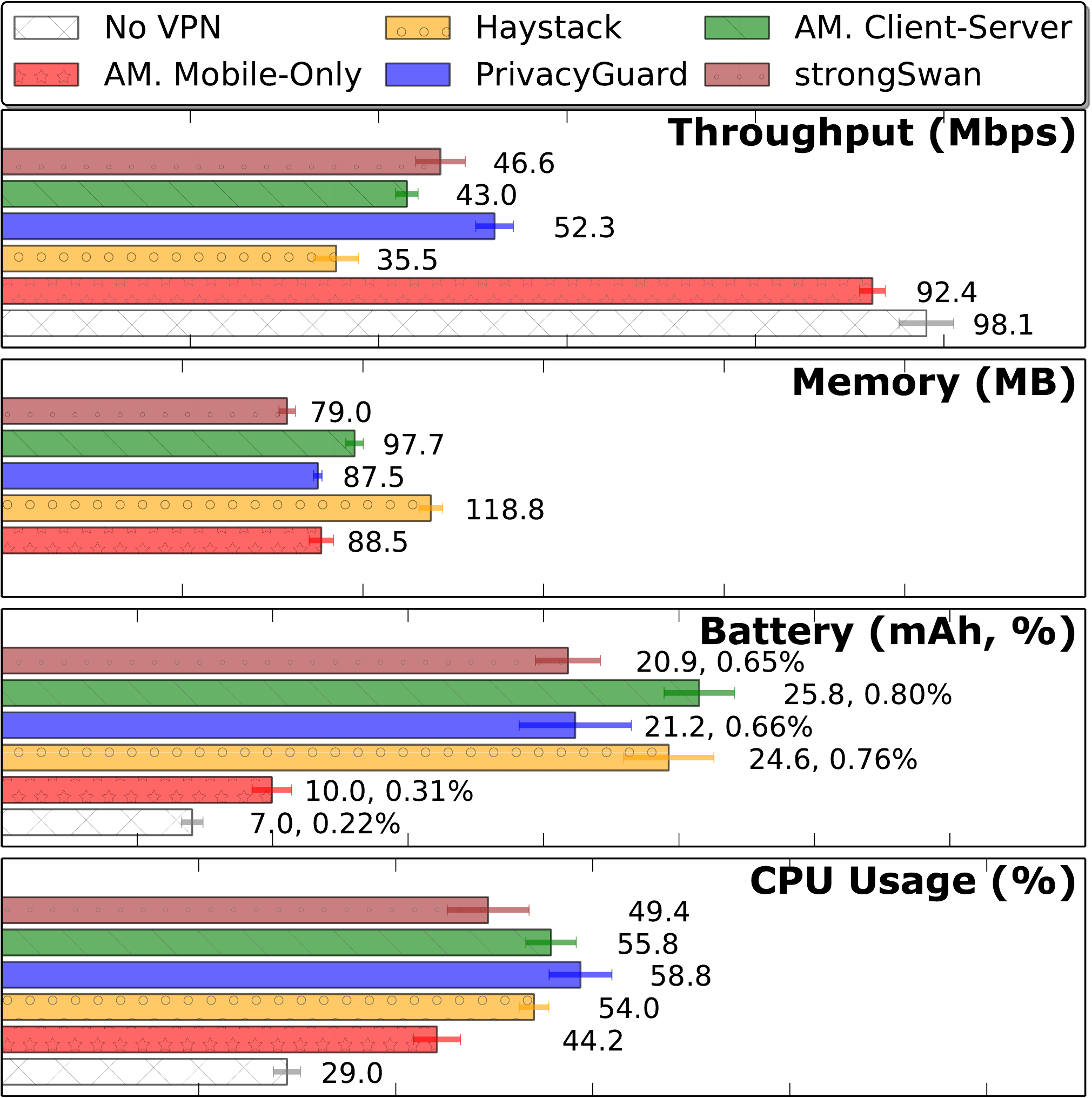}\label{fig:downPerf}}\hspace{5mm}
	\subfigure[Upload]{\includegraphics[height=5cm]{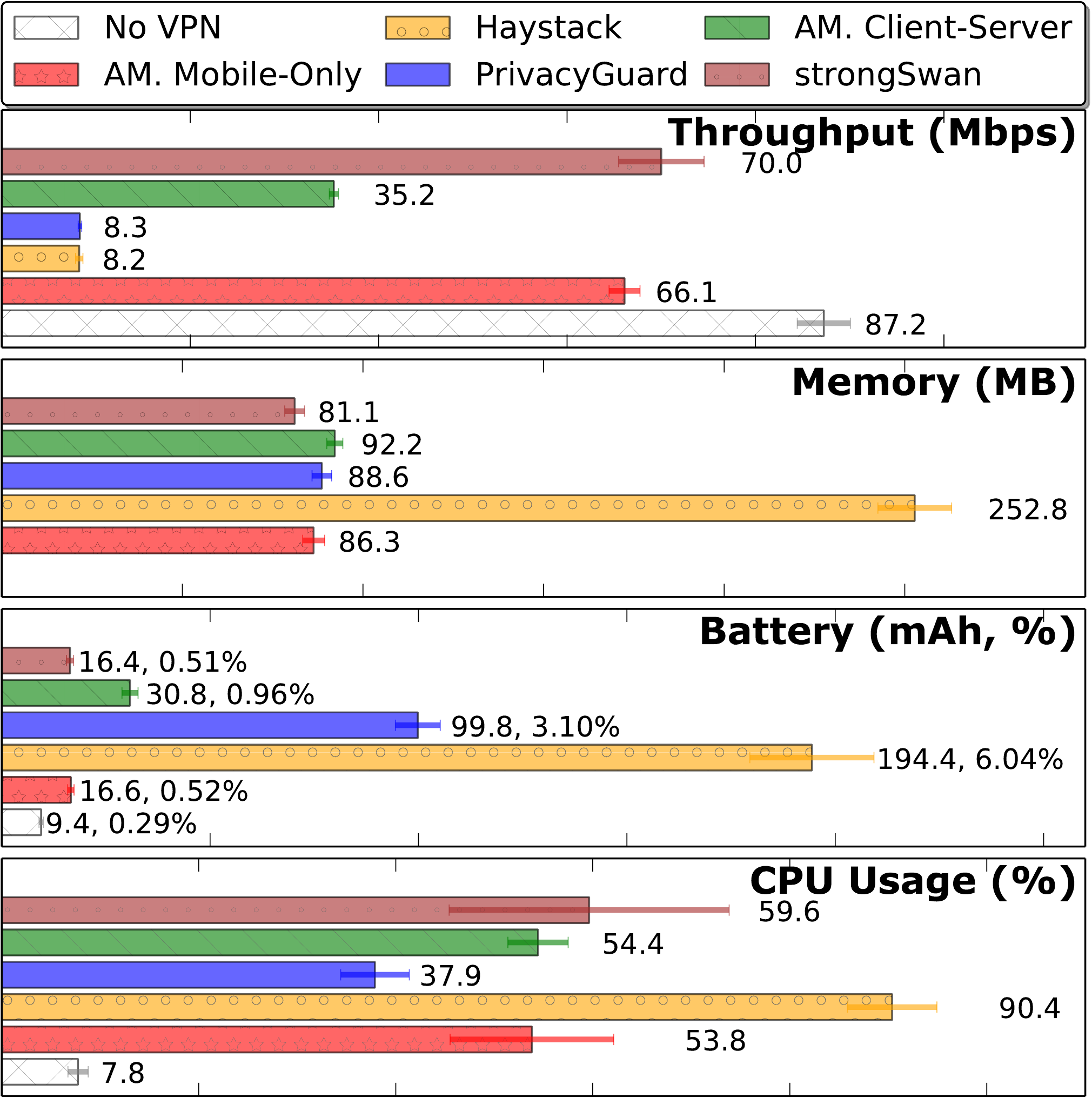}\label{fig:upPerf}}\hspace{2mm}
	\subfigure[4 Variations of AM. Mobile-Only Upload]{\hspace{3mm}\includegraphics[height=5cm]{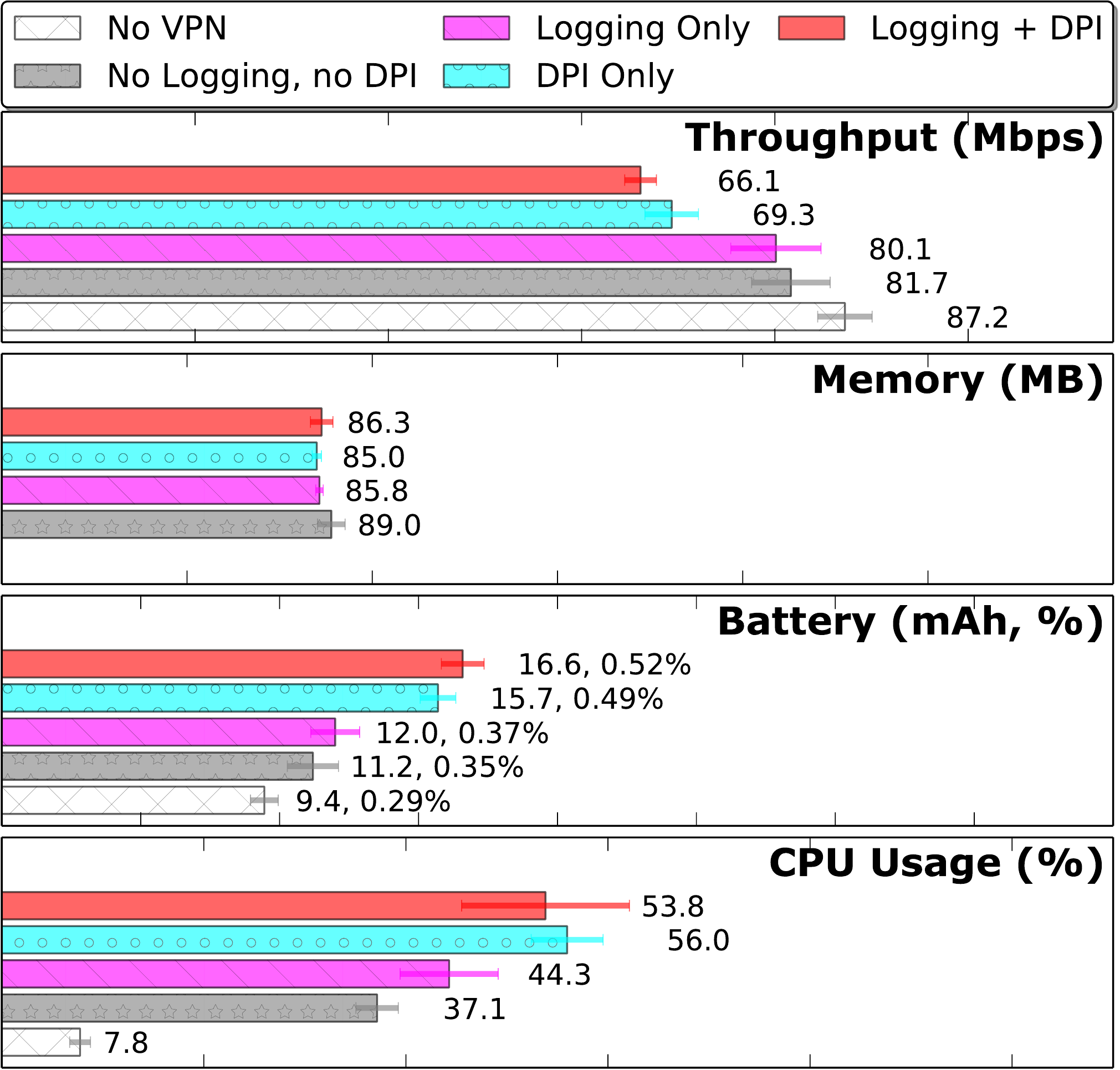}\label{fig:upPerfDetail}\hspace{3mm}}
	\caption{Performance of all VPN apps in Stress Test for a 500 MB file on Wi-Fi.  (``AM.'' stands for AntMonitor. ``AM. Mobile-Only'' stands for \anteater proposed in this paper. ``AM. Client-Server'' is only used as a baseline for comparison.)}
	\label{fig:performance}
\end{figure*}

\begin{figure*} [t!]
	\centering
	\subfigure[Download]{\includegraphics[height=4cm]{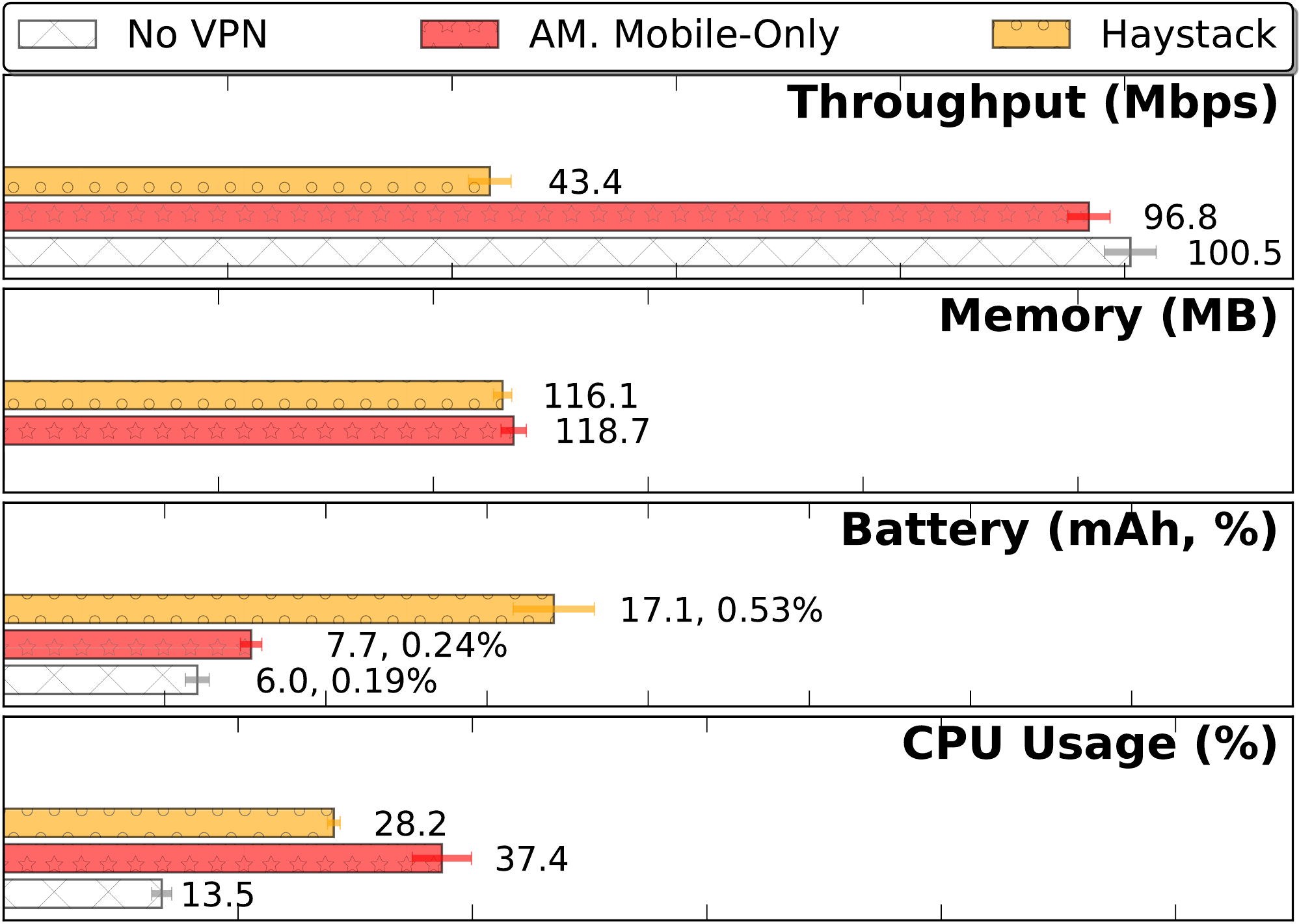}\label{fig:downPerf17}}\hspace{5mm}
	\subfigure[Upload]{\includegraphics[height=4cm]{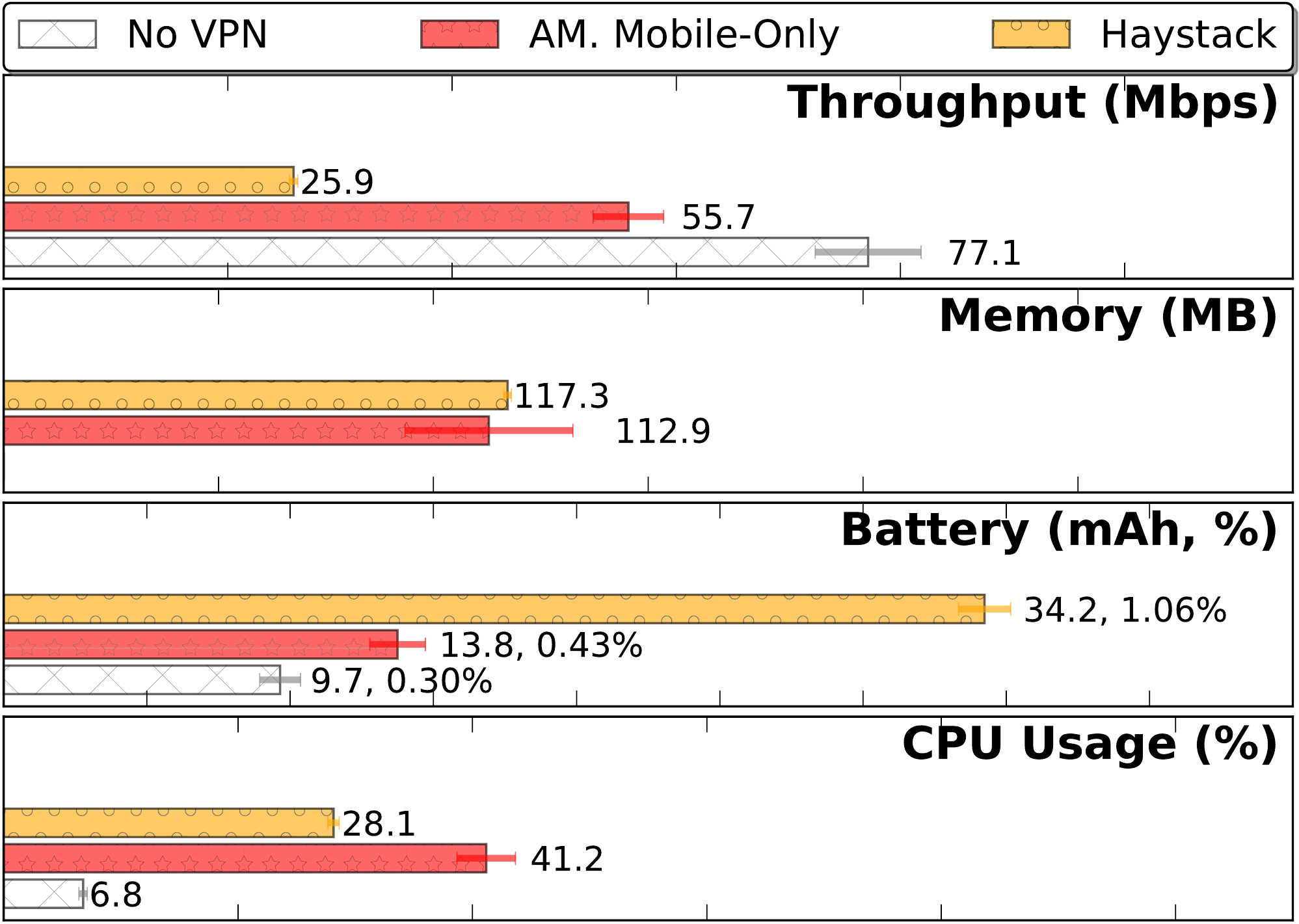}\label{fig:upPerf17}}\hspace{2mm}
	\subfigure[Performance during device idle time]{\hspace{3mm}\includegraphics[height=4cm]{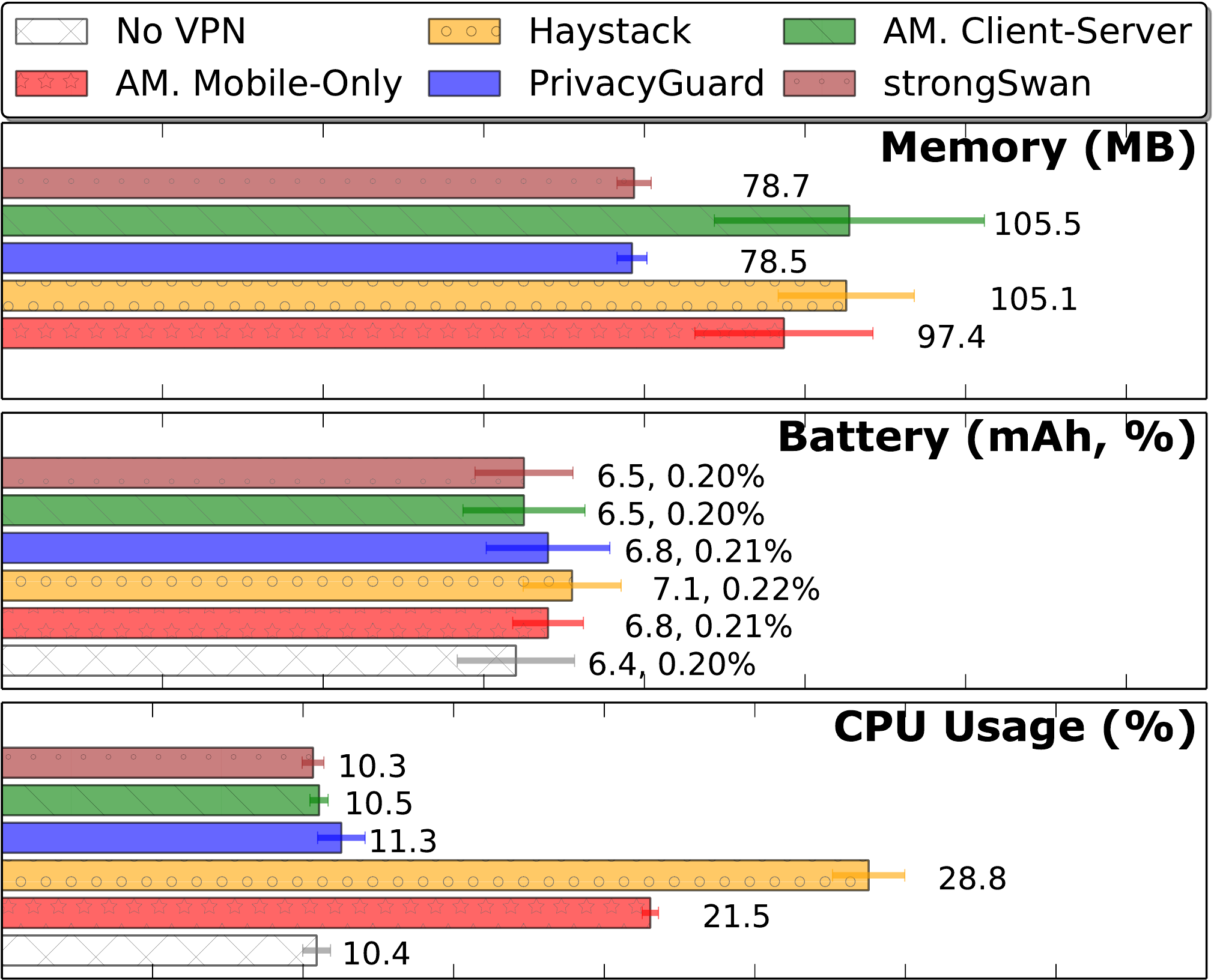}\label{fig:passivePerf}\hspace{3mm}}
	\caption{Performance of updated apps (in Feb. 2017) in Stress Test for a 500 MB file on Wi-Fi, and performance of all VPN apps during device idle time.}
	\label{fig:performance17}
\end{figure*}

{\bf Tool.} In order to evaluate  \anteater,  we built a custom app -- \evaluator. It transfers files and computes a number of performance metrics, including network throughput, CPU and memory usage, and power consumption. It helps us  tightly control the setup and compute metrics that are not available using off-the-shelf tools, such as {\em Speedtest}.

{\bf Scenarios.} We use \evaluator in two types of experiments. In Sec. \ref{sec:Stress Test}, {\em Stress Test} performs downloads and uploads of large files 
 so that \anteater has to continuously process packets. In Sec. \ref{sec:Passive test}, {\em Idle Test} considers an idling mobile device so that \anteater handles very few packets. In between the two extremes, we have also considered a {\em Typical Day Test}, which simulates user interaction with apps; however, it is omitted due to lack of space.

{\bf Baselines.} We report the performance of 
\anteater v0.0.1 and compare it to state-of-the-art baselines from  Sec. \ref{sec:VPN-related}: 
\begin{itemize}
\item \textvtt{Raw Device}: no VPN service running on the device; this is the ideal performance limit to compare against.
\item State-of-the-art mobile-only approaches:\\ \privacyGuard \cite{song2015privacyguard} v1.0 and \haystack \cite{razaghpanah2015haystack} v1.0.0.8. (We omit the testing of \tPacket \cite{tpacketcapture} since it was shown to have very poor performance in \cite{antmonitor15}.)
\item Client-server VPN approaches: industrial grade \swan VPN client v1.5.0 with server v5.2.1, and an \anteaterCS implementation based on \cite{antmonitor15}.\footnote{In the labels of the performance figures,  we refer to this baseline for comparison as ``AM. Client-Server,'' to distinguish it from the actual proposed \anteater,  referred to as ``AM. Mobile-Only''.} The VPN servers used by each app were hosted on the same machine.
\end{itemize}

{\bf Setup.} All experiments were performed on a Nexus 6, with Android 5.0, a Quad-Core 2.7 Ghz CPU, 3 GB RAM, and 3220 mAh battery. Nexus 6 has a built-in hardware sensor, Maxim MAX17050, that allows us to measure battery consumption accurately. Throughout the experiments, the device was unplugged from power, the screen remained on, and the battery was above 30\%. To minimize background traffic, we performed all experiments during late night hours in our lab to avoid interference,
we did not sign into Google on the device, and we kept only pre-installed apps and the apps being tested. 
Unless stated otherwise, the apps being tested had TLS interception disabled and the \client was logging full packets of all applications and inspecting all outgoing packets. In terms of versions, all tests were done with \anteaterMO v0.0.1 and \haystack v1.0.0.8, unless indicated otherwise (Sec. \ref{sec:SingleFlow}). VPN servers ran on a Linux machine with 48-Core 800 Mhz CPU, 512 GB RAM, 1 Gbit Internet; the Wi-Fi network was 2.4Ghz 802.11ac. %
 Each test case was repeated 10 times and we report the average.

\subsection{Stress Test} \label{sec:Stress Test}

\subsubsection{Large File Over a Single Flow} \label{sec:SingleFlow}

{\bf Setup.} For this set of experiments, we use \evaluator to perform downloads and uploads of a 500 MB file over a single TCP connection.
In the background, \evaluator periodically measures the following metrics:

{{\em A. Network Throughput:}} 
\evaluator reports the number of bytes transferred after the first 10 sec (to allow the TCP connection to reach its top speed) and the transfer duration. We use these numbers to calculate throughput.

{{\em B. Memory Usage:}} \evaluator uses the \textvtt{top} command to sample the Resident Set Size (RSS) value.

{{\em C. Battery Usage:}} \evaluator uses the APIs available with the hardware power sensor Maxim MAX17050  to compute the energy consumption during each test in mAh \cite{batteryAPI}. 

{{\em D. CPU Usage:}} \evaluator uses the \textvtt{top} command to measure the CPU usage. 

At the end of each experiment, \evaluator reports the calculated throughput and battery usage, and the average memory and CPU (considering the sum of CPU usage of \evaluator and the VPN app) usage.  

{\bf Results for AntMonitor v0.0.1 in Dec. 2015.} Fig. \ref{fig:downPerf} shows that the download throughput of \anteaterMO significantly outperforms all other approaches. It was able to achieve about {\em 94\% of the raw speed}, with throughput {\em 2x more} than \swan \& \privacyGuard and {\em 2.6x more} than \haystack. 
We further note that all VPN apps tested have similar memory, battery, and CPU usage.\footnote{Although \swan does not perform L3-L4 translation, it performs encryption and decryption, which results in about 5\% higher CPU usage than \anteaterMO.}  
Fig. \ref{fig:upPerf} reports the upload performance. \anteaterMO achieves {\em 76\% of the raw speed} while performing data logging and DPI.  Most significantly, its performance is {\em 8x faster} than both state-of-the-art mobile-only approaches.\footnote{The gains provided by the optimizations discussed in Sec. \ref{sec:opt} have a greater impact on upload speeds because both \privacyGuard and \haystack favor downstream traffic since the responses from the Internet are read as streams from sockets and the responses from applications are read from TUN packet-by-packet \cite{razaghpanah2015haystack}.}
\swan outperforms \anteater as expected since, unlike with incoming packets, \anteater performs DPI on each outgoing packet.
 Nevertheless,  Fig. \ref{fig:upPerfDetail} shows that \anteaterMO has the higher upload speed (and closest to the raw speed) if DPI is disabled. Fig. \ref{fig:upPerf} also shows that all VPN apps have similar memory and CPU usage, except for \haystack, which incurs significant overhead. Since the test took longer for the slower approaches, \privacyGuard and \haystack used significantly more battery.

In general, using any VPN service roughly doubles the CPU usage during peak network activity. Although the CPU usage of 38--90\% on Wi-Fi seems high, the maximum CPU usage on the quad-core Nexus 6 is 400\%. In summary, this set of experiments demonstrates that among all VPN approaches, for both downlink and uplink, \anteaterMO has the highest throughput while having similar or lower CPU, memory, and battery consumption.

{\bf Results for AntMonitor v0.1.5 in Feb. 2017.} Since the time we performed the above evaluation (in Dec. 2015), both \anteaterMO and \haystack (now renamed to ``Lumen'') were updated on GooglePlay, to versions 0.1.5 and 1.1.2, respectively. The main change for \anteaterMO from version 0.0.1 to 0.1.5 was improving (and making more complex) the GUI and disabling packet logging (in order to avoid dealing with the PIIs of the general public participating in the open-beta); the underlying design (including key optimizations, such as the use of two threads and minimum number of sockets, and the optimized reading/writing from the TUN) remains the same. The internal changes in Haystack/Lumen beyond the GUI are not available to us but the design of the system appears to remain the same, according to \cite{razaghpanah2016haystack}. To see how the updates affected the performance of both apps, we repeated a set of stress tests and we report these results separately in Fig. \ref{fig:performance17}(a-b). The network conditions have changed since the original tests were performed: \novpn throughput is now 100 Mbps and 78 Mbps on the downlink and uplink, respectively.  \haystack v1.1.2 has improved its download throughput to 43 Mbps, and \anteaterMO reached 96\% of the raw speed with a 96 Mbps throughput. The newer \haystack(Lumen) app also has improved its upload throughput to 26 Mbps. 
 The latest \anteaterMO has stayed in the 70\% range of the raw upload speed with a 56 Mbps throughput. This confirms that the design and optimization techniques applied to \anteaterMO still result in significant performance benefits, despite the added complexity of the updated GUI.

\subsubsection{Small Files Over Multiple Flows}

{\bf Setup.} To test the efficiency of \anteater's thread and socket allocation, we used \evaluator to create 16 threads, each downloading a 50MB file. During the test 
	\evaluator calculated the throughput of each flow and reported the average of all flows. %

{\bf Results.} The average speed of a flow (in Mbps) for each test case was the following: \textvtt{Raw Device}: 6.82, \anteaterMO: 6.57, \privacyGuard: 4.75, \swan: 3.73, \haystack: 3.18, and \anteaterCS: 3.06. Again, \anteaterMO came out on top, achieving 96\% of the raw speed.

\subsubsection{Impact of Logging and DPI} \label{sec:Impact of Log+DPI}
{\bf Setup.} To assess the overhead caused by Logging Data and Deep Packet Inspection (DPI), we performed 
the single-flow upload stress test on \anteaterMO with all four combinations of Logging on/off and DPI on/off. %

{\bf Results.}
First, Fig. \ref{fig:upPerfDetail} shows that logging does not have a significant impact on throughput. This is thanks to (i) the optimization of \anteaterMO 
 that uses only two threads for network I/O (see Sec. \ref{sec:opt}) and (ii) the fact that the data collection uses two threads for storage I/O.
These data logging threads do not significantly impact main network I/O threads on a quad-core Nexus 6 phone. Second, DPI is performed by one of the main network I/O threads and inflicts a 17\% slow-down on upload speed.  Although 17\% is a significant overhead, \anteaterMO is still able to reach over 60 Mbps speed, which is more than enough for mobile apps. In addition, DPI causes a 28\% and 33\% overhead on battery and CPU, respectively. However, the CPU usage still remains 1/8 of the total possible CPU available on the Nexus 6 (of 400\%), thus the overhead is acceptable. Finally, without logging and DPI, \anteaterMO achieves 94\% of the raw speed without VPN.
\subsubsection{Impact of TLS Proxy}

In order to be able to inspect encrypted traffic for privacy leaks, we implemented a TLS proxy, described in Sec. \ref{sec:leakage}. 

{\bf Setup.}  To evaluate the performance impact of this proxy, we used \evaluator to download a 500MB file from a secure server over HTTPS and compared the throughput  of \anteaterMO to that of the \textvtt{Raw Device}.

{\bf Results.} The average throughput (in Mbps) was 77.2 and 69.1 for the \textvtt{Raw Device} and \anteaterMO, respectively. As expected, the proxy causes a significant overhead since it uses an extra socket for each connection and performs one extra decryption and encryption operation per packet.

\subsection{Idle Test} \label{sec:Passive test}

{\bf Setup.} For this set of experiments, we kept the phone idle for 2 minutes with only background apps running. We used \evaluator to measure the battery and memory consumption of each VPN app. We also measured the aggregate CPU usage across all apps by summing the \textvtt{System} and \textvtt{User} \% CPU Usage provided by the \textvtt{top} command.


{\bf Results.} Fig. \ref{fig:passivePerf} shows that all apps tested create very little additional overhead when the device is in idle mode. Among the mobile-only approaches, \haystack and \anteaterMO used more CPU than \privacyGuard because both of them have threads to log packets while \privacyGuard does not. Similarly,  Logging also results in slightly higher memory usage for \haystack and \anteater. (Note that \swan does not log packets either, thus has lower CPU usage.) Finally, the overall memory usage of the VPN apps (\textasciitilde105 MB) is acceptable; many other popular apps, \eg Facebook, use as much as 200 MB of RAM.
\subsection{Metrics Computed Outside AntEvaluator} \label{sec:Other Metrics}

{\bf Latency.} We measured the latency of each VPN app by averaging over several pings to a nearby server (in the same city). In order of increasing delay, the apps rank as follows: \novpn: 3 ms, \swan: 4 ms, \haystack: 4 ms, \anteaterCS: 5 ms, \anteaterMO: 7 ms, and \pguard: 83 ms. Compared to client-server approaches, mobile-only approaches cannot forward ICMP packets; thus, we measure latency using TCP packets. The additional delay is due to the time required to create, send, and receive packets through TCP sockets. Compared to \haystack, \anteaterMO has a small additional latency as sending and receiving TCP packets involves two threads (one reads/writes the packet from/to TUN and one reads/writes the packet from/to the socket), whereas \haystack might have used a single thread (source code unavailable).

{{\bf String Parsing.}} The main heavy operation required in DPI is string parsing. During real traffic conditions, our native C implementation of Aho-Corasick has a maximum run time of 25 ms. When benchmarking as a standalone library (running on the Android main thread alone), our parsing time is <10 ms. For comparison, \haystack reports a 167 ms maximum run time for string parsing with Aho-Corasick. %

\section{Applications} \label{sec:applications}

\begin{figure}[t]
	\centering
	\includegraphics[width=0.45\textwidth]{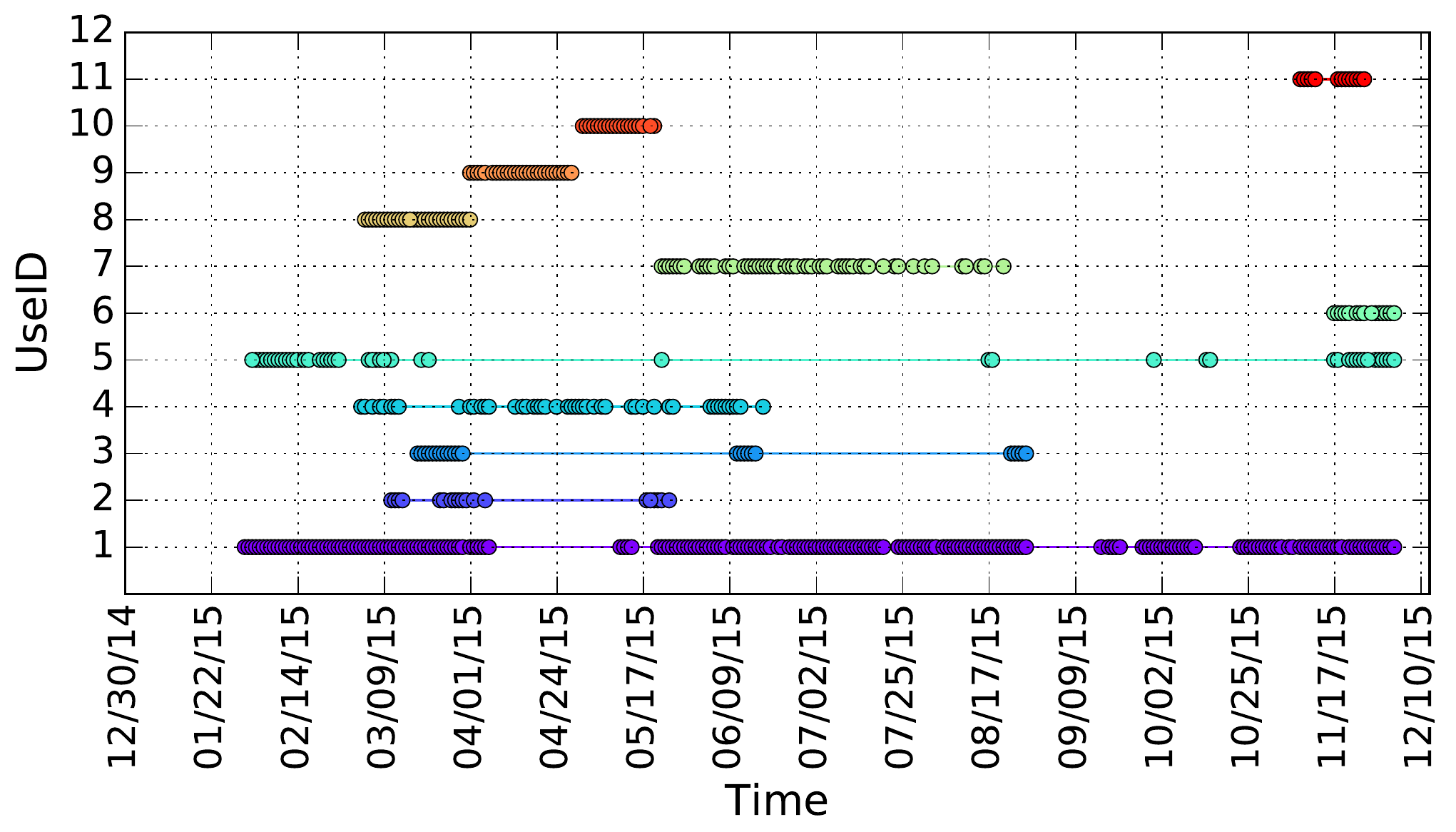} 
	\vspace{-5pt}
	\caption{Data logged daily by different users in the study. (We omitted some users, and we have included different devices belonging to the same individual (e.g. 7-11).)}
	\vspace{-10pt}
	\label{fig:activity}
\end{figure}

Because \anteater intercepts every packet in and out of the device, it is uniquely positioned to serve as a platform for enabling applications that build on top of its passive monitoring capability.  We showcase three such applications that may be of interest to operators and/or individual users: (i) privacy leaks detection, (ii) network performance monitoring, and (iii) traffic/user profiling. We report results from a pilot user study at our university campus: Fig. \ref{fig:activity} shows the daily activity of 11 (with re-installs) volunteers: these are members of our own research group, who used \anteater on their primary phones during the period  Feb. 5 -- Nov. 30, 2015 and uploaded their data to \logserver. This study was meant only as a proof of concept of the capabilities of \anteater and {\em not} as as representative large scale user study.

\subsection{Application I: Privacy Leaks}
\label{sec:leakage}

Mobile devices today have access to personally identifiable information (PII) and they routinely leak it through the network, often to third parties without the user's knowledge.  PII includes:  (i) mobile phone IDs, such as IMEI (which uniquely identifies a device within a mobile network), and Android Device ID; and (ii) information that can uniquely identify the user (such as phone number, email address, or even credit card) or other information (e.g. location, demographics). 
 Sometimes sending out PII is necessary for the operation of the phone (\eg  a device must identify itself to connect to a wireless network) or of the apps (\eg location must be obtained for location-based services). However, the leak may not serve the user (\eg going to advertisers or analytics companies) or may even be malicious. Leaks in plain text can be intercepted by listening third parties, \eg in public WiFi networks.
Although modern mobile platforms (Android, iOS, Windows) require that apps obtain permission before requesting access to certain resources, and  isolate apps (execution, memory, storage) from each other, this is not sufficient to prevent information from leaking out of the device, \eg due to interaction between apps \cite{felt2011permission}.  Even worse, users are unaware of how their data are used \cite{lin2012expectation}. 

\begin{table}
	\centering
	{\scriptsize
		\begin{tabular}{|@{}l@{}|@{}l@{}|@{}l@{}|}
			\hline
			{\bf App Name} &  {\bf Leak Type}  & {\bf \# Flows}\\
			\hline
			  VnExpress.net  & \multirow{2}{*}{ \parbox{2.2cm}{IMEI, Phone\#, Location, Email, DeviceID} }  & 33147 \\
				  &  & \\
			  Zing Mp3  & \xspace IMEI, DeviceID  & 14745 \\
			  Clean Master & \xspace DeviceID  & 1768 \\
			  WiFi Maps  & \xspace IMEI, Location  & 1582 \\
			  Relay for reddit  & \xspace Location  & 638 \\
			  System  &  \multirow{2}{*}{  \parbox{2.2cm}{IMEI, Location, DeviceID} }  & 301 \\
                                   &  & \\
			  Chrome  & \xspace Location  & 143 \\
			  ES File Expl.  & \xspace DeviceID  & 96 \\
			  MyFitnessPal  & \xspace Location  & 76 \\
			  DR Radio  & \xspace DeviceID  & 68 \\
			  Speedtest & \xspace Location  & 48 \\
			  DexKnows  & \xspace IMEI,Location  & 47 \\
			  Skype   & \xspace IMEI, DeviceID  & 42 \\
			  iWindsurf  & \xspace IMEI, Location  & 22 \\			  		
 			  WhatsApp &  Phone\#  & 20 \\	  			  			  
			$\cdots$ & $\cdots$ & $\cdots$ \\
			\hline
			  All & All & 52923 \\
			\hline
		\end{tabular}
		\quad
		\begin{tabular}{|@{}l@{}|l@{}|}
			\hline
			{\bf Domain Name} &   {\bf \# Flows}\\
			\hline
			eclick.vn. & 8760 \\
			api.mp3.zing.vn. & 7561 \\
			ksmobile.com. & 620 \\
			zaloapp.com. & 332 \\
			ngoisao.net. & 209 \\
			api.staircase3.com. & 150 \\
			openweathermap.org. & 47 \\
			adtima.vn. & 36 \\
			ads.adap.tv. & 31 \\
			apps.ad-x.co.uk. & 30 \\
			adkmob.com. & 27 \\
			whatsapp.net. & 27 \\
			mopub.com. & 25 \\
			duapps.com. & 25 \\
			api.dexknows.com. & 24 \\
			server.radio-fm.us. & 15 \\
			inmobi.com. & 14 \\
			$\cdots$ & $\cdots$ \\
			\hline
			All & 18020 \\
			\hline
		\end{tabular}
	}
	\caption{{\small Flows Leaking PII found in the collected data. (Note: the number of flows at the left (52923) is higher than at the right (18020) because we count a flow that leaks multiple PIIs multiple times.)}}
	\label{tab:leaks}
\end{table}

{\bf Privacy Leaks Detection and Prevention Module.} \label{sec:privacy-design}
We extended the basic \anteater functionality with an analysis module that performs real-time DPI.  The user can define strings that correspond to private information that should be prevented from leaking; see screenshot in Fig. \ref{fig:guiPI}. Before sending out a packet, the \client inspects it and searches for any of those strings. By default, if the string is found, \anteater hashes it (with a random string of the same length, so as not to alter the payload length) before sending the packet out, and asks the user what to do in the future for the given string/app combination, as shown in Fig. \ref{fig:guiNotif}. The user can choose to allow future packets to be sent out unaltered, block them, or keep hashing the sensitive string (so that the application has a good chance to continue working but without the string being leaked). The system remembers the action to take in the  future for the same app and ``leak,'' and it will no longer bother the user with notifications. The user may also look at the history of the leaks, shown in Fig. \ref{fig:guiHistory}. \anteater can provides both {\em real-time} detection and {\em prevention}, on the mobile-device (not at the server) thanks to its efficient implementation, while \haystack and \pguard do not achieve both; and \recon can achieve the same on the server, and could gracefully run on top of \anteater as well. Specifically, \anteater can inspect a packet for leaks in <10 ms (Sec. \ref{sec:Other Metrics}), and it was the first app to demonstrate this real-time capability in \cite{antmonitor-poster-mobicom15}.

{\bf Encrypted Traffic. } Since we require plain text in order to perform DPI, and much of the traffic is encrypted, we developed a TLS proxy that uses the SandroProxy library \cite{sandro-proxy}, also used by \pguard,  to intercept secure connections, decrypt the packets, and then re-encrypt them before sending them to their intended Internet hosts. This method works for most apps, but it cannot intercept traffic from highly sensitive apps, such as banking apps, that use certificate pinning. Due to the intrusive nature of intercepting TLS/SSL traffic, we allow users to disable this option.

\begin{figure}[t!]
\centering
\includegraphics[width=0.45\textwidth]{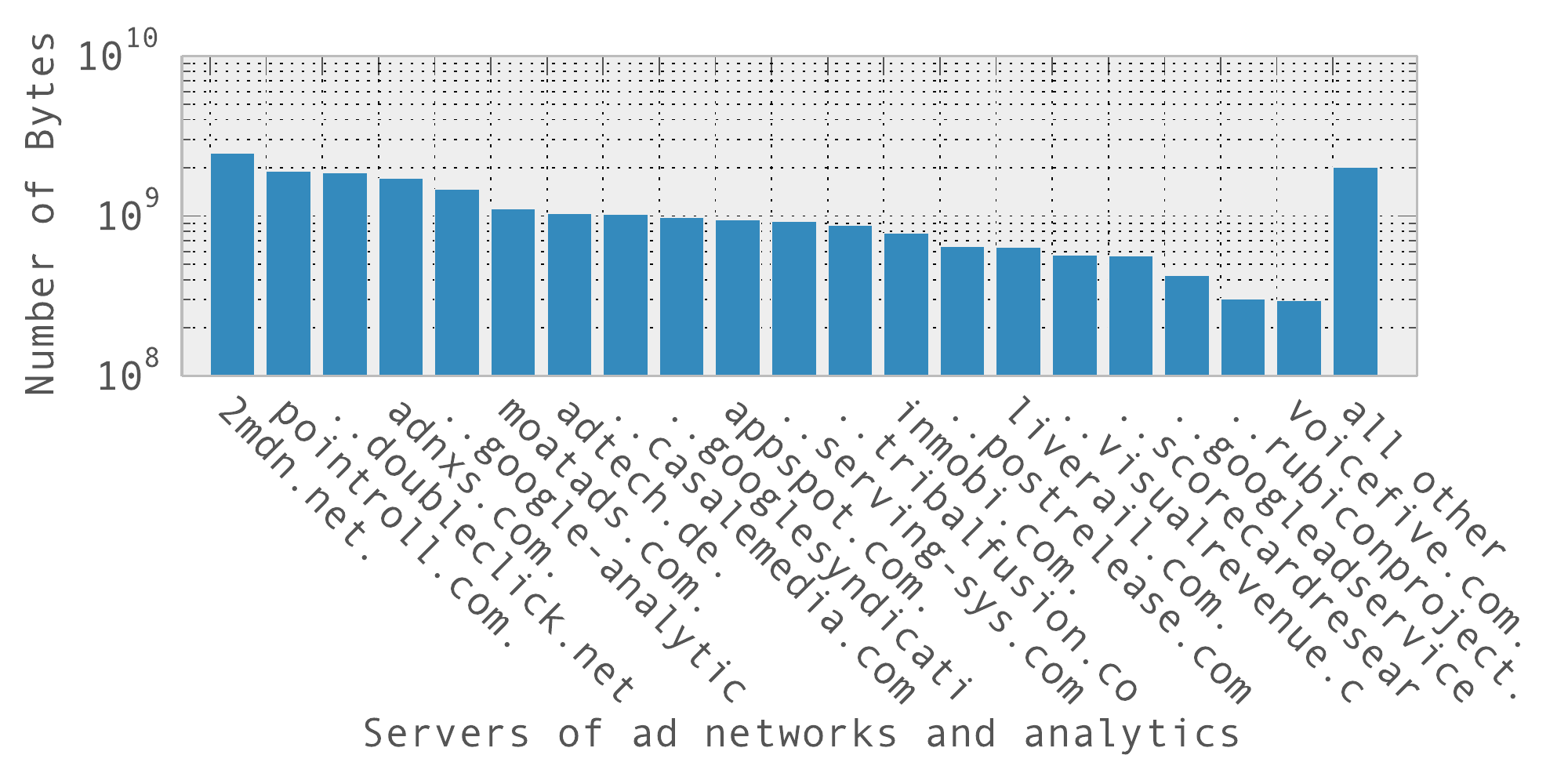}
%
\caption{Amount of traffic sent to ad \& analytics servers} 
\label{fig:adservers}
\end{figure}

{\bf Privacy Leaks Detected.} %
We analyzed the data collected from our user study, and found a large number of privacy leaks in plain text. Table \ref{tab:leaks} presents the apps and destination domains with the highest number of flows leaking. The worst offender in the list was the app \textvtt{VnExpress.net} that leaked five different types of PII up to 33,145 times towards the domain \textvtt{eclick.vn} -- an advertising network. The list of leaking apps includes very popular apps with tens of millions of downloads, such as Skype and WhatsApp, and the list of domains includes many  mobile ad networks (such as \textvtt{mopub, inmobi, adkmob, adtima}). 
Using the data collected by \anteater at the users' mobile devices, we calculated the amount of data that is transmitted from/to  ad networks, analytics and mediation services. %
 Fig. \ref{fig:adservers} shows that the amount of traffic transmitted towards such servers for the top 20 domains is in the order of GBs, consists of several domains, and tens of thousands of flows per domain. Fig. \ref{fig:guiVisual} visualizes the destinations for one device: it shows which apps leak information to which destinations. %

{\bf Discussion.} The inherent advantage of \anteater in this domain is that it runs on the device, which is a more attractive place for users to perform PII scrubbing than a middlebox, since sensitive data does not have to leave the device. Traffic encrypted at the application level can also be intercepted and inspected by the TLS proxy in \anteater, then re-encrypted, if the user chooses to do so (by agreeing to install a root certificate). One limitation of our current approach is that PII leakage is detected using string matching, which can be evaded by a more sophisticated  attacker, who can leak information across multiple packets or by encoding the text to be leaked. At its current state, our privacy  leaks detection is mostly useful for raising awareness of the magnitude, nature and cost of PII leaking, and can be useful in the case of an honest-but-curious adversary. For example, a legitimate app that leaks information to trackers is likely to stop doing so if users become aware and start uninstalling the app. The on-device visualization shown on Fig. \ref{fig:guiVisual} displays the destination (IP address, hostname/domain, and type of server) to which each app sends data for all active connections and it is updated in real-time. Another inherent advantage of running privacy leak detection on the device (as opposed to the middle server) is the access to contextual information available on the device (such as the app that generated the traffic), which can be used to detect more sophisticated PII leaks with machine learning.

\begin{table}[t!]
\centering
{\scriptsize
\begin{tabular}{| c@{} c@{} c@{} c@{} c@{} c@{} l l l l ||} 
 \hline
 Exp \# & \multicolumn{1}{|c@{}|}{1} & \multicolumn{1}{|c@{}|}{2} & \multicolumn{1}{|c@{}|}{3} & \multicolumn{1}{|c@{}|}{4} & \multicolumn{1}{|c@{}|}{5} & \multicolumn{1}{|c@{}|}{6} & \multicolumn{1}{|c@{}|}{7} \\ [0.5ex] 
 \hline\hline
 AM: W=1 & \multicolumn{1}{|r@{}|}{$30.00$} & \multicolumn{1}{|r@{}|}{$32.79$} &  \multicolumn{1}{|r@{}|}{$24.49$ } & \multicolumn{1}{|r@{}|}{$31.95$} & \multicolumn{1}{|r@{}|}{$34.78$} & \multicolumn{1}{|r@{}|}{$32.31$} & \multicolumn{1}{|r@{}|}{$31.02$}\\ 
 \hline
 AM: W=5 & \multicolumn{1}{|r@{}|}{$21.24$} & \multicolumn{1}{|r@{}|}{$29.26$} &   \multicolumn{1}{|r@{}|}{$22.83$ } & \multicolumn{1}{|r@{}|}{$27.01$} & \multicolumn{1}{|r@{}|}{$30.75$} & \multicolumn{1}{|r@{}|}{$26.84$} & \multicolumn{1}{|r@{}|}{$26.14$}  \\ 
 \hline
  Speedtest & \multicolumn{1}{|r@{}|}{$19.96$} & \multicolumn{1}{|r@{}|}{$28.42$} & \multicolumn{1}{|r@{}|}{$22.39$ } & \multicolumn{1}{|r@{}|}{$28.74$}& \multicolumn{1}{|r@{}|}{$31.66$} & \multicolumn{1}{|r@{}|}{$26.98$} & \multicolumn{1}{|r@{}|}{$27.22$}\\ 
 \hline
\end{tabular}
\caption{{\small {\bf Throughput (Download Mbps): Active (using Speedtest) vs Passive (using \anteater : AM) measurements}. First, we ran multiple Speedtests, with $5$ min gaps, from the same location, and we list the throughput mentioned by Speedtest. 
Second, we computed the (maximum) average throughput using \anteater logs, over a window of $1$ \& $5$ sec.
Our approach is close to Speedtest but does not incur any measurement overhead. For a fair comparison in this table, we passively monitored the Speedtest packets using AntMonitor. In the wild, throughput computations can be made by counting the bytes of actual traffic sent over time.
}}
\label{table:speedtest}
}
\end{table}

 \begin{table}[t!]

\centering
{\scriptsize
\begin{tabular}{| c@{} c@{} c@{} c@{} c@{} c@{} l l l l ||} 
 \hline
 Metric  & \multicolumn{1}{|c@{}|}{Data Overhead} & \multicolumn{1}{|c@{}|}{Memory} & \multicolumn{1}{|c@{}|}{CPU} & \multicolumn{1}{|c@{}|}{Battery} \\ [0.5ex] 
 \hline\hline
\speedtest & \multicolumn{1}{|c@{}|}{$50$ MB} & \multicolumn{1}{|r@{}|}{$116$ MB} &  \multicolumn{1}{|r@{}|}{$14.7\%$ } & \multicolumn{1}{|r@{}|}{$-0.5\%$}\\ 
 \hline
  \anteater & \multicolumn{1}{|c@{}|}{$0$ MB} & \multicolumn{1}{|r@{}|}{$134$} MB & \multicolumn{1}{|r@{}|}{$43.4\%$ } & \multicolumn{1}{|r@{}|}{$-0.7\%$}\\ 
 \hline
\end{tabular}
\caption{\small{\bf Resources Utilization} for \anteater vs. \speedtest .}
\label{table:comparison}
}
\end{table}

\subsection{Application II: Performance Measurements \label{sec:measurement}}
 %
 
In this section, we demonstrate how to use \anteater for monitoring network performance, which can benefit users (\eg, to manage their network access or cost) as well as operators (\eg,  to assess and improve their infrastructure \cite{cranorGigascope:03}).  Inherent advantages of \anteater in this application domain include the following:  (i) passive performance monitoring comes for free (without the need for active probing overhead) since \anteater touches every packet transmitted or received; (ii) the ability to correlate network-level metrics (\eg, WiFi and cellular speed) with other information available on the device (\eg, signal strength, type of network, location, time); (iii) the fine-granularity of measurements at the device, app, flow, or destination level.

{{\bf Performance Module.}} %
By design, \anteater intercepts every packet and is thus able to passively compute performance indicators of the TCP/IP layer, such as throughput and latency. In addition, it can monitor performance at other layers (\eg, the radio layer) and collect rich contextual information including but not limited to:  (i) timestamp; (ii) geolocation 
in a way that achieves a low energy footprint\footnote{\anteater listens to location updates by other applications passively and only supplements that with infrequent active requests.};
(iii)  network information (\eg, WiFi or Cellular), radio access technology (RAT), and detailed cellular network information per region 
(\eg, LTE parameters, frequency bands);
(iv) received signal strength (RSS), and
(v) throughput and latency measurements per app and overall.
For the purposes of this paper, we append RSS, geolocation, RAT, and network connectivity information in PCAPNG files based on network connectivity/location events posted by the OS. The analysis of the collected data 
is done offline at \logserver.

We utilize our module to {\em passively} compute the smartphone's throughput and we compare it to a state-of-the-art {\em active} monitoring tool (\speedtest). Table \ref{table:speedtest}  shows that the values are very close, but our passive approach does not incur any data overhead. Resources usage by these two methods is shown in Table \ref{table:comparison}. For a fair comparison in Table \ref{table:speedtest}, we passively monitored the Speedtest packets using \anteater. In the wild, throughput computations can be made by counting the number of bytes of actual traffic sent over time.

{{\bf Example Measurements.}}  In the rest of the section, we report measurements collected on our campus (for one month Nov.7-Dec. 7, 2015 and among approx. 10 people in our group) including: reference signal received power (RSRP) for LTE network and device throughput (both WiFi and cellular).

 %
%

 \begin{figure*}[t!]
\centering
\subfigure[Crowdsourcing]{\includegraphics[scale=0.1]{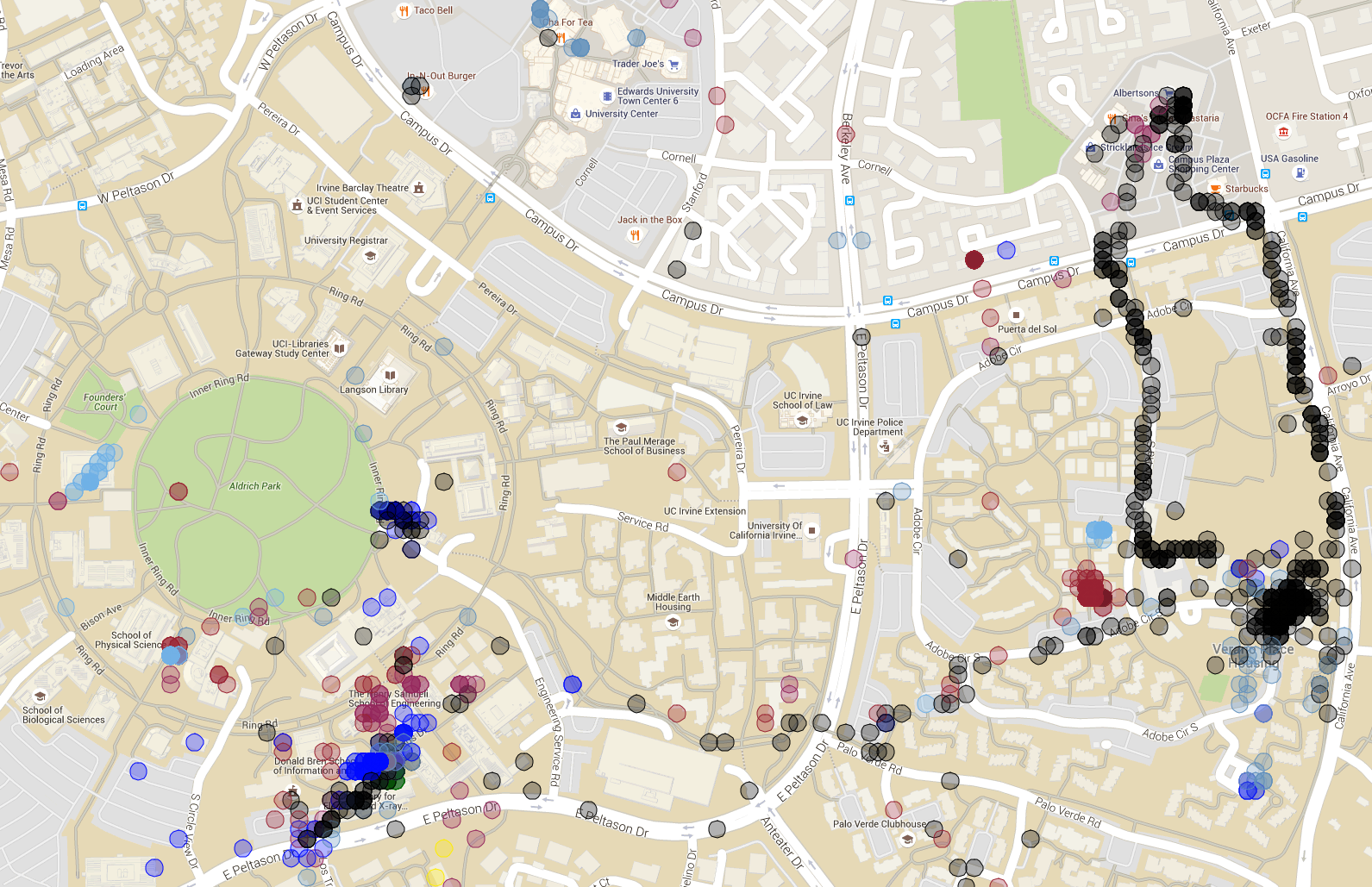}} \hspace{5pt}
\subfigure[T-Mobile LTE Signal Strenth, Wi-Fi \& Cellular Throughput Maps]{\includegraphics[scale=0.12]{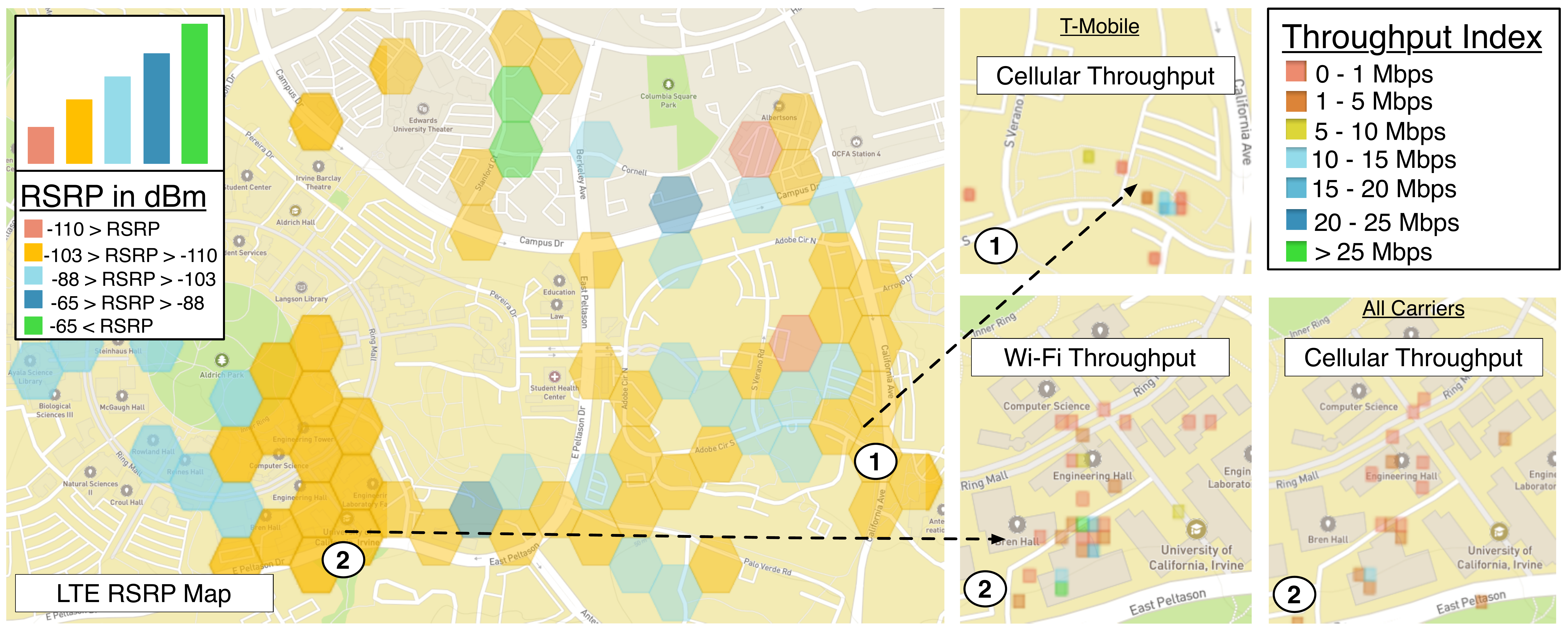}\label{fig:maps_all}}
\caption{\small Performance maps from the university campus. (a) Monitor a large area with only a few users. (b) The university has many areas with moderate to poor LTE coverage and many spatial variations. However, low RSRP (location 1) does not necessarily mean low cellular throughput (for the same carrier). \label{fig:maps}} 
\end{figure*}

\begin{figure}[t!]
\centering
\subfigure[Week Typical Day]{\includegraphics[scale=0.205]{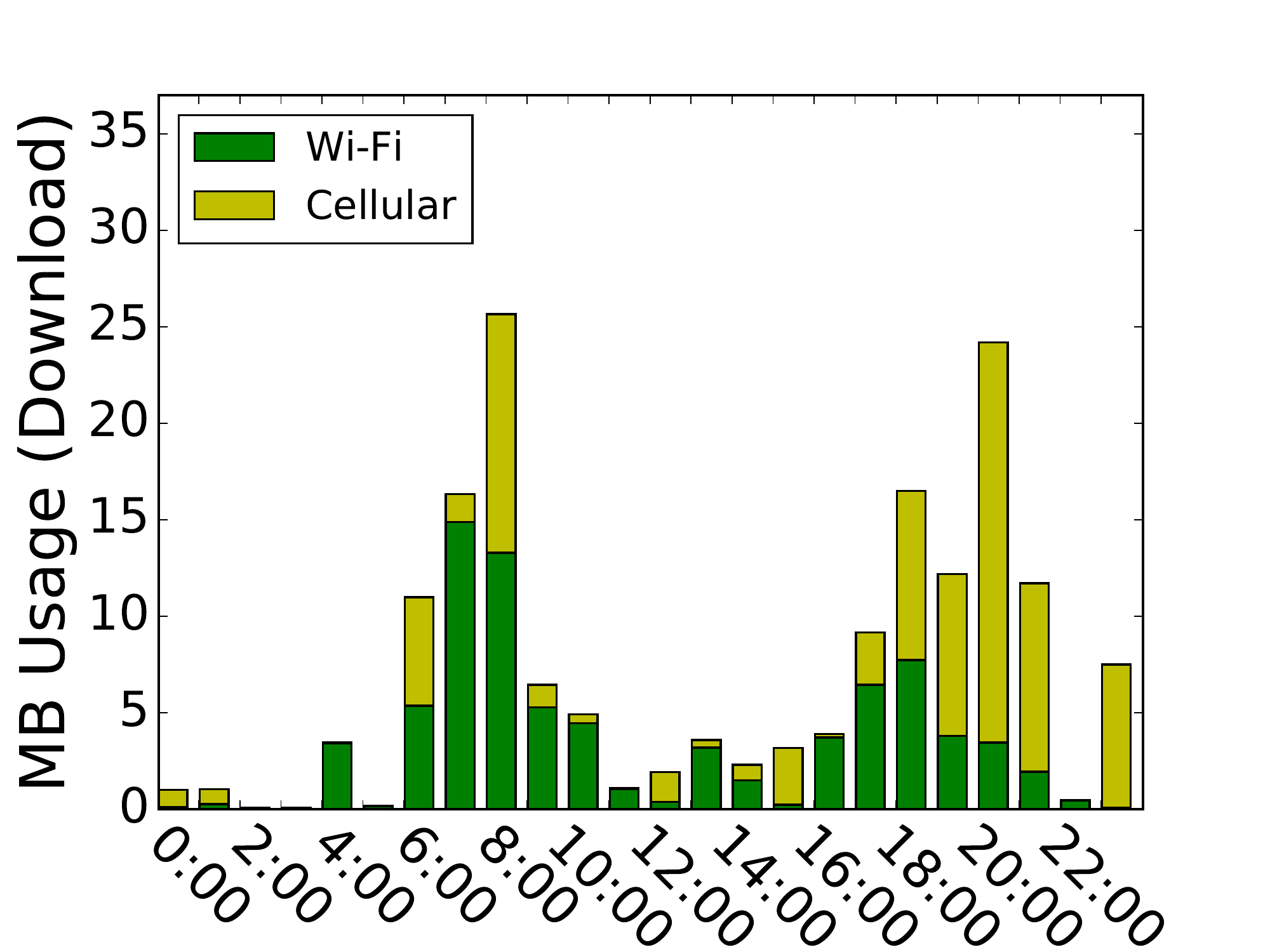}\label{fig:sampleUser_typWeekDay}}
\subfigure[Weekend Typical Day]{\includegraphics[scale=0.205]{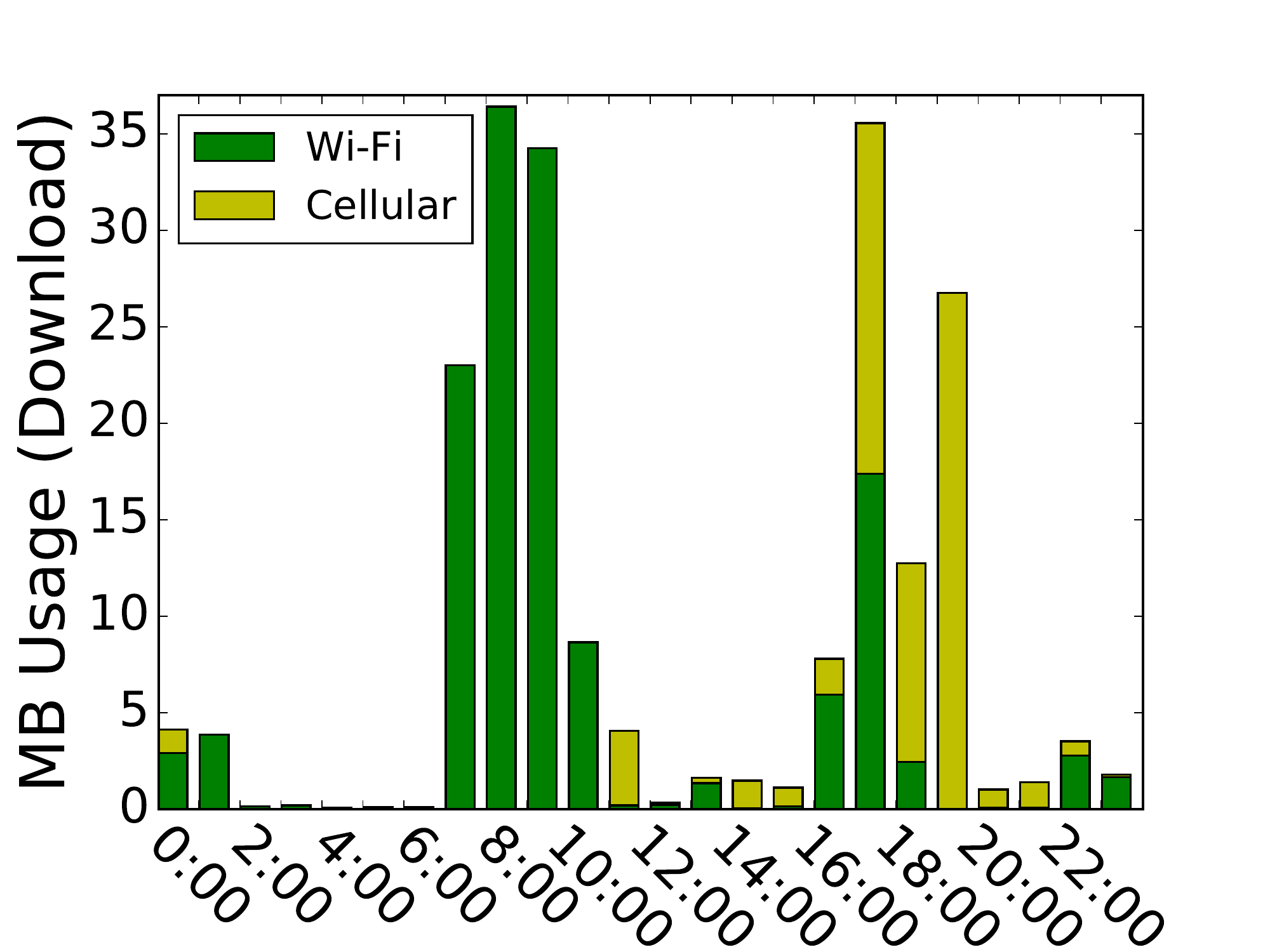}\label{fig:sampleUser_typWeekendDay}}
\caption{{\small {\bf Single user's point of view:} \#MB downloaded, averaged over all (a) week or (b) weekend days, for one month, for one user. }\label{fig:sampleUser_typicalDay}}
\end{figure}

Fig. \ref{fig:sampleUser_typicalDay} shows the data (MB) used by one user throughout a typical day (\ie, averaged over all  week or weekend days) per network (WiFi or Cellular). One can see the breakdown of the traffic  into WiFi and cellular, the daily pattern and the difference between week and weekend. Such statistics are currently reported by mobile devices, but typically at a very coarse granularity (\eg, total amount of data left for this month). In contrast, \anteater can report data at fine granularity (\eg, per flow, per app, per location, over time, etc.) and can enable a vast number of monitoring applications, troubleshooting, and SDN operations.

In ongoing work, we plan to incorporate this module of \anteater to the open beta app, and crowdsource performance measurements, which can  provide a comprehensive view of network-wide performance. As an illustrative example, Fig. \ref{fig:maps} depicts 5 such performance maps we built for our university campus, reporting signal strength and QoS parameters  per cellular carrier and WiFi. Fig.~\ref{fig:maps}(a) depicts the location of 5 users reported during the one month period (Nov. 7-Dec. 7, 2015): only with a few users, we are able to cover a large area of the campus. Network performance varies throughout the day,
but  thanks to minimal battery overhead,   \anteater can monitor it continuously and passively, allowing us to  collect time-series of network performance. %
Fig. \ref{fig:maps}(b) depicts the LTE RSRP for one cellular provider: LTE reception is poor on many areas and has large spatial variation. Fig. \ref{fig:maps}(b) also reports the average throughput of WiFi and cellular networks on campus and compares it to LTE RSRP. This information can, for example, be used to inform decisions about switching among, as well as to better provision these networks. Interestingly,  in Fig.~\ref{fig:maps}(b),  location 1 has low RSRP but high  cellular throughput. %
 With the increasing complexity of cellular networks, %
 \anteater can help  by simultaneously assessing multiple layers of network performance.

\subsection{Application III: Learning Network-Level Behavior}
\label{sec:learning}

\anteater passively monitors all network traffic in and out of the device and can use TCP/IP header features and other contextual information to learn profiles at different levels of granularity, including: per device, app, flow, destination, etc. These can enable anomaly detection, user profiling, market research, traffic differentiation, etc.
A key, inherent advantage of \anteater, in this application domain,  is that it has access to rich contextual information available on the device (\eg location, time, background/foreground apps, application that generated the packets\footnote{  
For example, although not the first to classify traffic based on packet headers, \anteater has an advantage because it operates on the device: for every packet \anteater intercepts, it can identify the app it belongs to with 99\% accuracy, and append it to the PCAPNG file. This provides ground truth, which is hard to get.}
) that can be used to train accurate classification models.
We demonstrate two examples: (i) flow classification to mobile apps they belong to, see Sec. \ref{sec:learning-flows} and (ii) learning user profiles from the apps they use, see Sec. \ref{sec:learning-users}. These can be useful building blocks for anomaly detection (on the device), traffic differentiation (by the ISP), market research, etc.  Training and classification can be performed on the device (Log module in the \client)  and/or at the \logserver (where data is contributed by multiple devices). In the rest of the section, we report results from the latter. 

\begin{figure}[t!]
\centering
\includegraphics[width=0.45\textwidth]{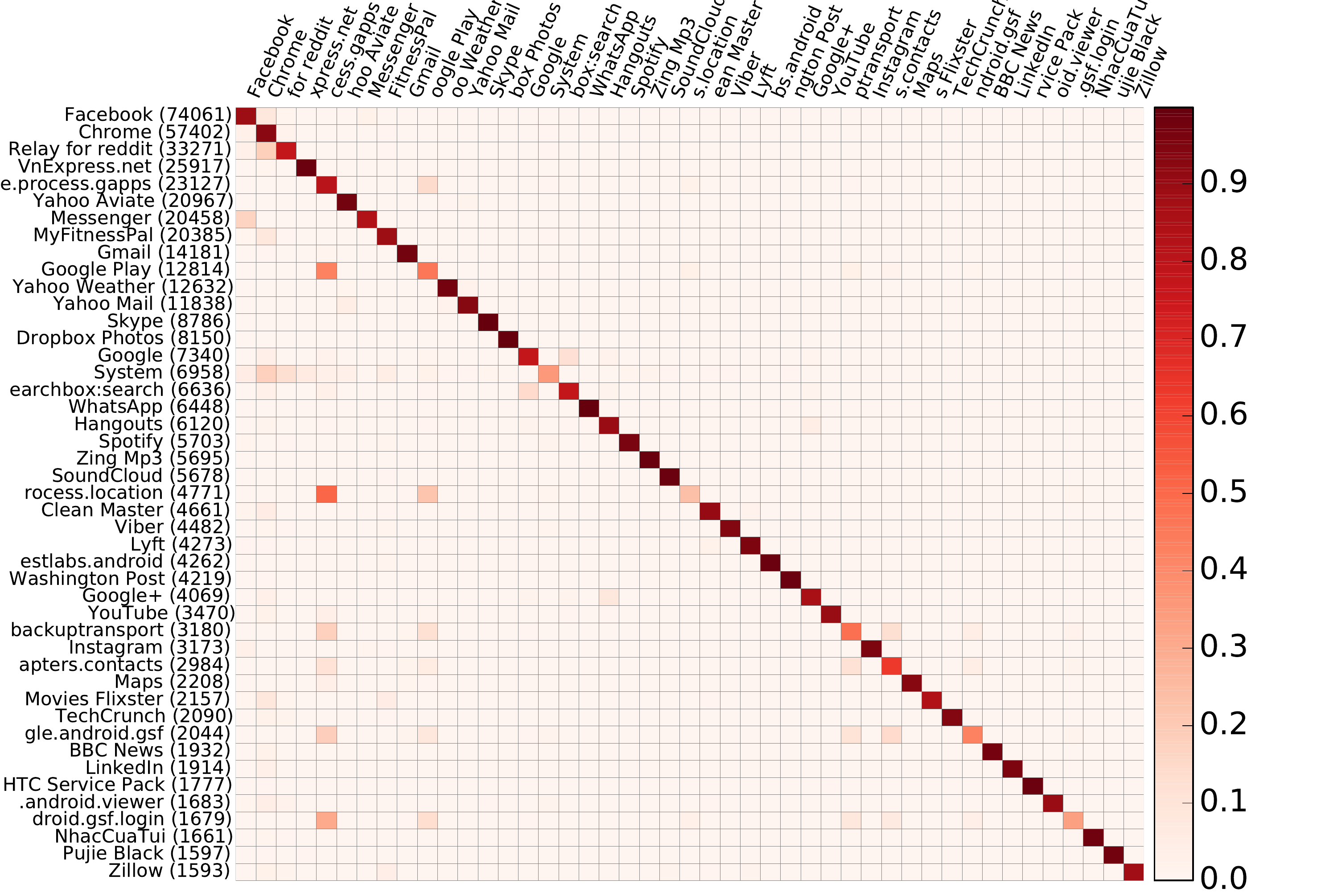}
\caption{ \small{Classification of Flows to Mobile Apps based on TCP/IP features. Normalized  confusion matrix for all features. Parentheses show \#flows used in training \& testing.} \label{fig:confusion_matrix} }
\end{figure}

\subsubsection{Application classification\label{sec:learning-flows}}
We use  packet headers  collected in the user study, together with app names that generated the traffic, to train models and classify flows to mobile apps. We used supervised learning to build a multi-class model that classifies network flows to apps.
 For each flow, we extracted 66 flow features from layer-3 and layer-4 headers (\ie payload and packet size statistics, burstiness, packet inter-arrival times, TCP flags, flow statistics, IP features)  on the upstream \& downstream directions. %
 We compared different ML models and selected the Random Forest, which performed best. We used a 10-fold cross-validation  and kept the same proportions of  apps in the testing and training. 
When we used all features together, the F-1 score was up to 78\%. Flow classification for an individual user further increases the classification performance, with the F-1 score ranging between 75\%-93\%; this is expected, since the number of apps per user is smaller (ranges between 25-70). As a baseline for comparison, random uniform and random proportional classification yield  F-1 scores of only 1.5\% and 5.8\% respectively. \textvtt{Meddle} reports a 64.1\% precision score in classifying flows for the 92 most popular Android applications by using payload features (Host and User-Agent) \cite{meddle}. 
For a dataset of millions of applications, the state-of-the-art approach of AppPrint, that also requires HTTP header data, achieves 81\% flow-set coverage with 91\% precision.   Fig. \ref{fig:confusion_matrix} zooms in the results for the top 45 apps and shows which apps are correctly classified (diagonal entries), while the few errors (non-zero entries off-diagonal) misclassify similar apps to each other (\eg, Facebook to FB  Messenger, or Google related apps).  
The fact that  using off-the-shelf learning tools and only features from TCP/IP headers, \anteater can classify applications better than state-of-the-art approaches with  access to payload, is due to its inherent advantage of having access to accurate ground truth and user behavior at a large scale.

\begin{figure}
\centering
\includegraphics[width=0.3\textwidth]{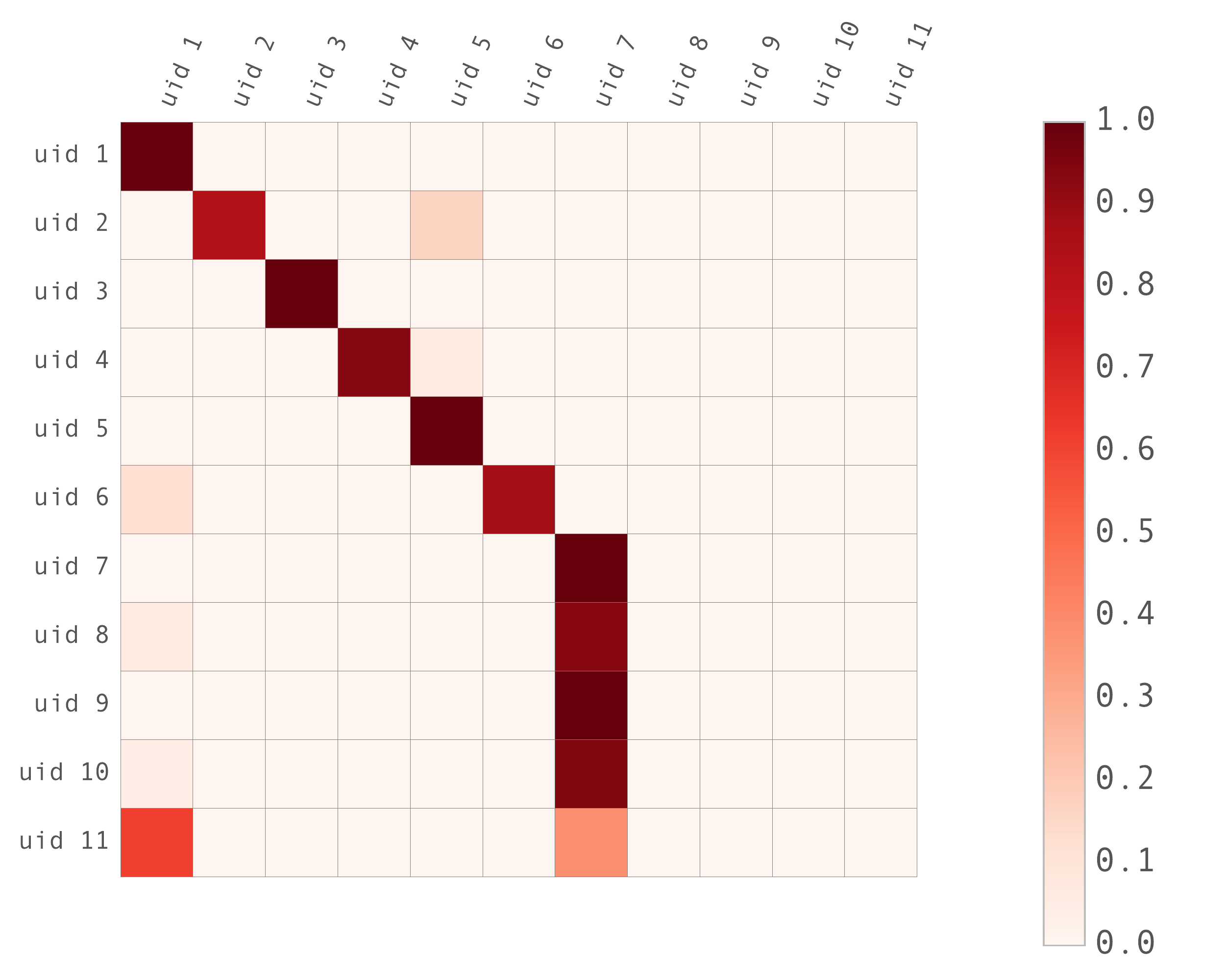}
\caption{Supervised Classification of Users based on the normalized volume of their mobile apps activity. } 
\label{fig:user_classification}
\end{figure}

\subsubsection{User Profiling\label{sec:learning-users}}

As a representative example, we asked the following question: can the users in our study group (see Fig. \ref{fig:activity}) can be distinguished from each based on their  app activity? We model each user as the vector of normalized activity volume per app, in one day. %
One interesting fact in our dataset was that certain users have re-installed \anteater during the study and appear with different user ids. For example, users 7-11 in Fig. \ref{fig:activity} are different devices used by the same person over different time periods. First, we use supervised learning in which we include users 1-7 in the training dataset and users 1-11 in testing. Fig. \ref{fig:user_classification} shows the confusion matrix: users 1-7 are correctly classified to themselves, while users 8-10 are classified  mostly as user 7, which is also correct. Interestingly, user 11 is classified as user 1, which also makes sense: during that period the 2 users (students in our group) were working on the same  deadline and were using their phones for running similar apps for testing and performance evaluation.  %
 These results are clearly preliminary but demonstrate \anteater's  potential for user profiling and anomaly detection, a direction we plan to explore in the future.

\section{Conclusion and Future Directions} \label{sec:conclusion}

The focus of this paper was on the \anteaterMO system for on-device (as opposed to client-server) passive mobile network monitoring. 
 Although VPN-based approaches have been used before, \anteater's optimized design minimizes the use of resources  and significantly outperforms previous state-of-the-art  mobile-only approaches, namely \pguard \cite{song2015privacyguard} and \haystack \cite{razaghpanah2015haystack}: it achieves 2x and 8x faster (down and uplink) speeds and close to the raw no-VPN throughput, while using 2--12x less energy.  This significant performance benefit is crucial for the successful adoption of \anteater (users are unlikely to install apps that slow down their phones or drain their battery) and for enabling some real-time applications (\eg, privacy leak prevention was not possible before on a mobile-only design).  

The applications discussed were meant to showcase the inherent advantage of the \anteater system to support them, and they are a standalone topic on their own right. Our pilot deployment at our university campus was also limited to the primary phones of the members of our research group, for alpha testing purposes and as a proof-of-concept; it was not meant as a large scale user study.  Despite these limitations, we demonstrated that \anteater is  a powerful tool for end-users to understand the behavior (at the network, application, and device level) of their device, detect and block privacy leaks,   understand where data is transmitted to and correlate patterns; and for operators to correlated network performance measurements at various layers.    

Our ongoing and future work includes the following. On the systems side, we have packaged the \anteaterMO functionality as an SDK that can be used by application developers and researchers inside their own apps. We are currently working with 4 such partners, and we eventually plan to open-source the SDK for the research community. 
Our goal is to increase our base of end-users, both directly (open beta on GooglePlay) and indirectly (through the SDK and third party apps). In terms of applications, we are working on automating the privacy leaks detection using machine learning and exploiting the access to ground truth available on the device; and on distinguishing legitimate use of PIIs vs privacy leaks. More generally, we envision that the \anteater SDK can provide a crowdsourcing platform for collecting data and enabling data transparency and performance optimization, with a competitive advantage being  its optimized design and superior performance, combined with the advantages that stem from running on the device as opposed to a middle server.



\end{document}